\documentclass[12pt]{article}
\pdfoutput=1
\usepackage{color,amsmath,amssymb,exscale,psfrag,epsfig}
\usepackage{cite,color,url}
\usepackage{pdfpages}
\usepackage{graphicx}
\usepackage{slashed}
\usepackage[utf8]{inputenc}
\usepackage[T1]{fontenc}
\usepackage{xcolor}
\usepackage{accents}
\usepackage{centernot}
\usepackage{ulem}
\usepackage{enumerate}
\usepackage{subfig}
\usepackage{float}
\usepackage{placeins}
\usepackage{floatpag}

\usepackage[colorlinks=true
,urlcolor=blue
,anchorcolor=blue
,citecolor=blue
,filecolor=blue
,linkcolor=blue
,menucolor=blue
,linktocpage=true
,pdfproducer=medialab
]{hyperref}
\input epsf
\usepackage{lineno}

\def\bea{\begin{eqnarray}}
\def\eea{\end{eqnarray}}

\definecolor{nicered}{rgb}{0.7,0.1,0.1}
\definecolor{nicegreen}{rgb}{0.1,0.5,0.1}
\hypersetup{colorlinks,citecolor= nicegreen,linkcolor= nicered}

\newcommand{\change}[1]{{\color{black} #1}}

\def\be{\begin{equation}}
\def\te{\end{equation}}
\def\ee{\end{equation}}
\def\ba{\begin{eqnarray}}
\def\bea{\begin{eqnarray}}
\def\nn{\nonumber}
\def\tea{\end{eqnarray}}
\def\ea{\end{eqnarray}}
\def\eea{\end{eqnarray}}

\def\d{\delta}

\def\bfra{\begin{frame}}
\def\efra{\end{frame}}
\def\al#1\fal{\begin{align}#1\end{align}}
\def\bfra#1\efra{\begin{frame}#1\end{frame}}

\catcode`\@=11
\def\lsim{\mathrel{\mathpalette\@versim<}}
\def\gsim{\mathrel{\mathpalette\@versim>}}
\def\@versim#1#2{\vcenter{\offinterlineskip
\ialign{$\m@th#1\hfil##\hfil$\crcr#2\crcr\sim\crcr } }}
\catcode`\@=12
\catcode`@=12
\begin{document}
\thispagestyle{empty}
\begin{flushright}
ICAS 045/19
\end{flushright}
\vspace{0.1in}
\begin{center}
{\Large \bf Topic Model for four-top at the LHC} \\
\vspace{0.2in}
{\bf Ezequiel Alvarez$^{(a)\dagger}$,
Federico Lamagna$^{(a,b)\ddag}$,
Manuel Szewc$^{(a)\diamond}$
}
\vspace{0.2in} \\
{\sl $^{(a)}$ International Center for Advanced Studies (ICAS), UNSAM, Campus Miguelete\\
25 de Mayo y Francia, (1650) Buenos Aires, Argentina }
\\[1ex]
{\sl $^{(b)}$ Centro At\'omico Bariloche, Instituto Balseiro and CONICET\\
Av.\ Bustillo 9500, 8400, S.\ C.\ de Bariloche, Argentina}
\end{center}
\vspace{0.1in}

\begin{abstract}
	We study the implementation of a Topic Model algorithm in four-top searches at the LHC as a test-probe of a not ideal system for applying this technique.  We study this Topic Model behavior as its different hypotheses such as mutual reducibility and equal distribution in all samples shift from true.  The four-top final state at the LHC is not only relevant because it does not fulfill these conditions, but also because it is a difficult and inefficient system to reconstruct and current Monte Carlo modeling of signal and backgrounds suffers from non-negligible uncertainties. We implement this Topic Model algorithm in the Same-Sign lepton channel where S/B is of order one and all backgrounds cannot have more than two b-jets at parton level.  We define different mixtures according to the number of b-jets and we use the total number of jets to demix.  Since only the background has an anchor bin, we find that we can reconstruct the background in the signal region independently of Monte Carlo.  We propose to use this information to tune the Monte Carlo in the signal region and then compare signal prediction with data.  We also explore Machine Learning techniques applied to this Topic Model algorithm and find slight improvements as well as potential roads to investigate.  Although our findings indicate that still with the full LHC run 3 data the implementation would be challenging, we pursue through this work to find ways to reduce the impact of Monte Carlo simulations in four-top searches at the LHC.

\end{abstract}

\vspace*{2mm}
\noindent {\footnotesize E-mail:
{\tt 
$\dagger$ \href{mailto:sequi@unsam.edu.ar}{sequi@unsam.edu.ar},
$\ddag$ \href{mailto:federico.lamagna@cab.cnea.gov.ar}{federico.lamagna@cab.cnea.gov.ar},
$\diamond$ \href{mailto:mszewc@unsam.edu.ar}{mszewc@unsam.edu.ar}
}}

\newpage
\section{Introduction}
\label{section:1}

The LHC is a very successful machine that is achieving amazing results such as the discovery of the Higgs boson \cite{higgs_disc1,higgs_disc2}, the detailed study of the top quark \cite{topprop1,topprop2} and Higgs boson \cite{higgsprop1,higgsprop2} properties, and the current exploration of the physics in the TeV scale as no other available experiment.  The latter is pushing the frontiers of our knowledge by testing many of the available new theories.  The LHC is currently designed to collect in the next twenty years approximately an order of magnitude more data than what has been collected so far.  To enhance the LHC discovery scope beyond this increase in luminosity, it is crucial that the community commits to explore and find new observables and different analyses which can further probe the Standard Model (SM) beyond current tests.  For these reasons, we investigate in this article the perspectives for a new analysis on the four-top final state based on Topic Model techniques. 

Four-top is arguably among the last benchmarks of Standard Model Physics at the LHC together with $t\bar t h$ \cite{atlastth,cmstth} and $hh$ \cite{micco2019higgs,atlashhprod,cmshhprod} final states. In particular, four-top is especially sensitive to light New Physics (NP) which couples predominantly to the top quark \cite{alvarez2019fourtop,Alvarez:2016nrz}. There exists also in the literature many works that point out four-top as a sensitive channel to test heavy NP and/or Effective Field Theory effects \cite{Englert_2019,4tops1,4tops2,4tops3,4tops4,lucien}.  However, since it is a very populated final state and with many different possible channels, its reconstruction is inefficient and therefore its measurement \cite{Sirunyan:2019wxt, Sirunyan:2019nxl} is carried on mainly through signal regions which are compared to Monte Carlo (MC) predictions.  Although MC-predictions have reached a complex level of development, including Next to Leading Order (NLO) matrix level calculations for four-top production \cite{zaro}, there are still many measurements of four-top signal and backgrounds which require renormalizations to data in control regions which still need further understanding.  Because of this, we investigate in this work a direction to reduce the impact of MC simulations and tuning in the extraction of physical quantities in the four-top final state.   It is worth stressing at this point that, even with all the new Machine Learning and Topic Model techniques, it is not possible to avoid a dependence on MC simulations in extracting absolute physical quantities from the LHC data.  We pursue the goal of reducing the impact of these simulations by replacing some of the required simulation and/or calibrations with different techniques.

\change{Topic modeling is a subject of natural language processing and Machine Learning that concerns the study of statistical models to recover abstract topics that occur in a corpus of documents. One of the seminal works in this subject defines the Latent Dirichlet Allocation (LDA) \cite{Blei:2012:PTM:2133806.2133826} technique to determine and extract topics from a corpus of documents. It considers topics as probability distributions over words, and resorts to different strategies for recovering these distributions from the observed word distributions in the documents. Some works build over LDA, as for instance the Dynamic Topic Model \cite{dynamicTM}, that considers ordered documents within the corpus, the Correlated Topic Model \cite{correlatedTM}, that considers correlation among the topics, the Online LDA \cite{onlineLDA} that allows for a streaming of the documents in the training process, or the Decontamination of Mutual Contamination Models \cite{katz2017decontamination}, that tackles the demixing of mixed membership models, among others. Some of these sets of tools have recently found their way into jet physics, where now documents are the distributions over some jet observable, and words are each of the bins in these distributions\cite{Dillon:2019cqt}. The underlying topics then cease to be abstract notions, but are meaningful distributions, for example signal and background for a particular process. For the simple case in which there are two observed distributions, and two underlying distributions to be recovered, a data-driven implementation of a mixture membership topic modeling has been applied to differentiate quark from gluon jets \cite{Metodiev:2018ftz,Komiske2018}. Along this work we investigate this particular framework applied to four-top physics.    We refer to this technique as the {\it Topic Model Demixer algorithm}, or {\it Demixer algorithm} for short \cite{Metodiev:2018ftz}.}

We investigate in this article the four-top final state and its backgrounds as two specific topics which are mixed in different proportions in two defined samples according to the number of jets that are b-tagged in the final state.  We explore within this framework the perspectives for measuring four-top physical properties, combining the information that can be extracted from the Demixer algorithm with MC simulations and techniques.  

This work is divided as follows.  In Section \ref{section:2} we present the Demixer algorithm and investigate its behavior in cases where its hypotheses are not fully satisfied.  In Section \ref{section:3} we apply the Demixer algorithm to four-top production and its main backgrounds $t\bar t W$ and $t\bar t H$, where we show how to recover important physical properties and distributions without relying on MC simulations.  In Section \ref{section:4} we discuss improvements and alternative strategies that could be performed on the algorithm to enhance the result in this work.  We discuss in Section \ref{section:4} a possible strategy to tune a MC generator using the information extracted from the Demixer algorithm.   We summarize our conclusions in Section \ref{section:5}.  In Appendix \ref{appendix} we essay Machine Learning techniques on the Demixer algorithm to improve the results in describing the signal and background distribution and purity fractions.

\section{Demixer algorithm} 
\label{section:2}

In this section we give a brief review on the Demixer algorithm. We first describe the algorithm along with its basic features, which allows to recover two underlying distributions and their fractions from a pair of mixed samples. We then discuss under which conditions the algorithm works properly, and we analyze more realistic cases which relax some of these conditions.  We study how the algorithm can still be used to recover sensible topics, as long as the withdrawal of the hypotheses is tractable.

\subsection{Demixer algorithm in the ideal case}
Along this work we consider the case of two samples $M_1$ and $M_2$ that are a mixture, in different proportions, of two underlying sources which we call signal and background.  In the following paragraphs we summarize the basic layout of the problem for the ideal case, further details can be found elsewhere \cite{Metodiev:2018ftz,katz2017decontamination}.  

Assuming that some features of the elements in samples $M_{1,2}$ can be described by a given observable $x$, we can define a probability distribution of these elements in each sample as
\be
p_{M_i}(x) = f_i\, p_S(x) + (1-f_i)\, p_B(x).
\label{samples}
\te 
Here $p_S(x)$ and $p_B(x)$ are the two underlying distributions in $x$ of signal and background, respectively, and $f_i$ is the fraction of signal events in sample $M_i$.  These are the unknowns that one would like to recover, or {\it demix}, from the original samples $M_1$ and $M_2$.  Observe that one of the important assumptions is that $p_S(x)$ and $p_B(x)$ are the same for both samples.

In order to demix these samples we perform a maximal subtraction through the definition of reducibility factors:
\be
\kappa_{ij} \equiv \kappa\left(M_i|M_j\right) \equiv \textrm{min}\left( \frac{p_{M_i}(x)}{p_{M_j}(x)}\right)
\label{kappaij}
\te 
We maximally subtract sample $M_2$ from sample $M_1$ and normalize it in order to define the distribution of the reconstructed signal topic $T_S$ over $x$ as:
\be
p_{T_S}(x) = \frac{p_{M_1}(x) - \kappa_{12}\, p_{M_2}(x) }{1-\kappa_{12}}.
\te
In a similar way, we can obtain the reconstructed background topic $T_B$ distribution in $x$.  Without loss of generality, we suppose that sample $M_1$ has a larger fraction of signal events than sample $M_2$ ($f_1 > f_2$), and therefore $T_S$ is the one matching the signal distribution, whereas topic $T_B$ would match the background distribution.  

For the demixer to work properly, besides having different fractions $f_i$ in Eq.~\ref{samples}, we need to have anchor bins. That is, having at least two points (or bins, in the discrete case) $x_S$, $x_B$ such that
\al
p_S(x_S)=0,&\quad p_B(x_S)\neq 0,\quad \nn \\
p_S(x_B)\neq 0,&\quad p_B(x_B)= 0. \nn 
\fal 
Defining the irreducibility factors for the underlying distributions $\kappa(B|S)$ and $\kappa(S|B)$ by replacing $M_{i,j}$ by $S$ and $B$ in Eq.~\ref{kappaij}, this implies that the reducibility factors $\kappa(B|S) = \kappa(S|B) = 0$.  This hypothesis is usually called \textit{mutual irreducibility}. When this condition is guaranteed, topics reconstructed from maximally subtraction samples between themselves will match underlying signal and background distributions, and the fractions $f_{1,2}$ can be recovered from inverting the following equations: 
\be
\kappa_{12} = \frac{1-f_1}{1-f_2}, \quad\quad \kappa_{21} = \frac{f_2}{f_1}
\te 
If there is no mutual irreducibility, sample demixing still leads to relevant quantities. For instance, the reconstructed topic for the more signal-like sample leads to the background-subtracted signal distribution:
\be
p_{T_S}(x) = p_{S|B} (x) = \frac{p_S(x) - \kappa_{SB} \, p_B(x)}{1-\kappa_{SB}}.
\label{substracted}
\te 
And the analogous is valid for the other sample, by swapping $B$ and $S$. If there is extra input on the size of these reducibility factors, by either theoretical principles, or by some given estimation, then the two equations can be solved for the pure distributions. \change{In this sense, $\kappa_{SB}$ and $\kappa_{BS}$ can be thought as hyperparameters, since prior information on them would provide a better determination of the underlying topic distributions.} 

\subsection{Demixer algorithm beyond the ideal case}

In the above paragraphs we have shown the basic features in extracting two topics from two samples that contain different mixtures of these topics in an ideal case.  In the real case scenario the procedure is highly different and many factors may affect the conclusions.  For instance, since the data has uncertainties, the minimum in Eq.~\ref{kappaij} may not correspond to the true minimum, or one of the underlying distributions may not have an anchor bin, or the signal or backgrounds distributions may be different in $M_1$ and $M_2$, among others.  In the following paragraphs we perform a brief study on how some of these real case scenario factors may affect the extraction of topics and fractions.  In this overview we neglect experimental factors such as statistic and systematic uncertainties and we focus in studying cases where some of the hypotheses needed for the Demixer algorithm procedure is not satisfied.  A more exhaustive study scrutinizing the effect of all these factors lies beyond the scope of this work.  Nevertheless, it would be useful for further understanding the reliability of the Demixer algorithm in particle physics where not only the ideal case is always far from reality, but also because in particle physics the topics (and their fractions) are usually not abstract distributions, but instead have physical meaning.

Let us consider cases in which the mixture samples stray from the optimal conditions for demixing. In order to keep track of these conditions, we define two basic hypotheses as:
\begin{itemize}
\item {\bf H1:} Mutual irreducibility. Both underlying signal and background have anchor bins. 
\item {\bf H2:} Same underlying distributions. The two samples are sums of the same signal and background distributions, differing only on the fractions.
\end{itemize}

In many real case scenarios H1 is relaxed either because mutual irreducibility is a priori unknown, or because it is well known that one of the underlying distributions does not have an anchor bin.  Relaxing H1 means that the reconstructed topics now match background-subtracted-signal distribution and vice-versa, as in Eq (\ref{substracted}). However, if one could know the reducibility factors $\kappa_{SB}$ and $\kappa_{BS}$, inverting this system of equations would give the true underlying distributions.   In such cases one option is to resort to either theoretical or experimental arguments, or simulations in order to estimate the size of the reducibility factors.  In the same direction, one could justify that any of the reducibility factors is zero --or negligible-- and extract useful information, even though the other is unknown.  In fact, as shown in Ref.~\cite{Komiske2018} or in Eq.~\ref{substracted}, this means that one can reconstruct the distribution that has an anchor bin exactly without recurring to additional information.

In order to quantitatively discuss deviations from H2 we need to quantify the difference between probability distributions.  There are many options to quantify the difference between two curves, however given that these two curves are probability distributions we adopt the Hellinger distance \cite{distances} which is defined as 
\be
d^2_H(P_A,P_B) \equiv \frac12 \int \! dx \, \left(\sqrt{P_A(x)} - \sqrt{P_B(x)}\right)^2  = 1 - \int \! dx \, \sqrt{P_A(x) P_B(x)}. 
\te
This distance has the property that through the use of a square root it provides a relative enhancement in importance to regions with smaller values. The purpose of this is to be more sensitive to differences in the small probability regions, where the anchor bins are located.  In any case, it is worth mentioning that we have verified that using other notions of distance such as $L^1$ or $L^2$ does not yield qualitative changes to our conclusions.

With this notion of distance, we can measure the Demixer performance when the hypothesis of same underlying distributions is not fulfilled. We define $\d_{S,B}$ to quantify that either signal or background distributions can be different in each sample as
\al
\d_S &\equiv d_H(P_{S_1},P_{S_2}) \nn \\ 
\d_B &\equiv d_H(P_{B_1},P_{B_2}). \nn
\fal 
Here the sub-index 1 or 2 refers to the underlying distributions in samples $M_1$ and $M_2$, respectively.  To determine the performance of the algorithm, we can measure how good signal and background reconstruction is by defining the distances
\al
\Delta_{S_i} &\equiv d_H(P_{T_S},P_{S_i}) \nn \\
\Delta_{B_i} &\equiv d_H(P_{T_B},P_{B_i}). \nn 
\fal 
Since sample $M_1$ is the one with larger fraction of signal, and sample $M_2$ the one with larger fraction of background, we define for practicality $\Delta_S=\Delta_{S_1}$ and $\Delta_B=\Delta_{B_2}$.

In order to keep the same notion of mutual irreducibility when in presence of four underlying samples $P_{S_{1,2}}(x)$ and $P_{B_{1,2}}(x)$, we consider the deviation between underlying samples to be generated by multiplicative noise, that is
\be
P_{S_2}(x) = P_{S_1}(x) \left(1+\xi_S(x)\right) \nn
\te 
with a noise function $\xi_{S}$ satisfying
\be
\int \! dx \, P_{S_1}(x) \, \xi_S(x) = 0  \label{xinorm}
\te 
to keep $P_{S_2}(x)$ normalized.  For the background we have analogous relationships exchanging $S \rightarrow B$ and $1 \leftrightarrow 2$.  In this way $P_{S_1}(x)$ and $P_{B_2}(x)$ would be the reference distributions for signal and background, respectively.

To better explore the Demixer algorithm conditions and their subsequent relaxation we split the space of possible combinations into a few cases. For H1 we have either 
\begin{itemize}
\item Case a) Mutual irreducibility
\item Case b) Only one anchor bin
\item Case c) No anchor bins
\end{itemize}
If H2 holds, in case b) the algorithm still recovers correctly the distribution that has the anchor bin.  If in addition the reducibility factors are known, in all three cases the demixer works in recovering underlying distributions. For H2 we distinguish three cases:
\begin{itemize}
	\item Case 0) $P_{S_1}(x)=P_{S_2}(x),\quad P_{B_1}(x) = P_{B_2}(x)$
\item Case 1) $P_{S_1}(x) \neq P_{S_2}(x),\quad P_{B_1}(x) = P_{B_2}(x)$
\item Case 2) $P_{S_1}(x) \neq P_{S_2}(x),\quad P_{B_1}(x) \neq P_{B_2}(x)$
\end{itemize}
Case 0) is where H2 holds. For case 1) there is a mirrored case that occurs by switching between $S$ and $B$ labels. 
Case 2) is the most general one, and can be particularized to the other cases for either $\d_{B} \to 0, \d_{S} \to 0$ or both.

We can see how the demixer works when H2 does not hold. In this case the factors $\kappa_{ij}$ get modified due to the difference between underlying signal and background functions
\be
\kappa_{12} = \left(\frac{1-f_1}{1-f_2}\right)\frac{P_{B_1}(x_S)}{P_{B_2}(x_S)},\quad \kappa_{21} = \left(\frac{f_2}{f_1}\right)\frac{P_{S_2}(x_B)}{P_{S_1}(x_B)} . \label{kappasmodified}
\te 
As we parameterize the difference between functions as a multiplicative noise, we get
\be
\kappa_{12} = \left(\frac{1-f_1}{1-f_2}\right)\left(1+\xi_B^0 \right),\quad \kappa_{21} = \left(\frac{f_2}{f_1}\right)\left(1+\xi_S^0\right).
\te
Here $\xi_B^0$ and $\xi_S^0$ are the values of the noise functions at the anchor bins, which should not be confused with the functions. Performing the maximal subtraction on sample $M_1$ yields --assuming $f_1 > f_2$-- the signal-topic
\begin{eqnarray}
	P_{T_S}(x) &=& P_{S_1}(x) \frac{(1-f_2)f_1 - f_2 (1-f_1)(1+\xi_S(x))(1+\xi_B^0) }{1-f_2 -(1-f_1) \xi_B^0} \nn \\ && + P_{B_2}(x) \frac{(1-f_1)(1-f_2)\left(\xi_B(x)-\xi_B^0\right)}{1-f_2 -(1-f_1) \xi_B^0}
\end{eqnarray}
The background-topic is found by replacing $f_1 \to 1-f_2$, $f_2 \to 1-f_1$, and flipping $B$ and $S$ labels. We can first consider the simpler case 1), which corresponds to $p_{B_1}(x) \equiv p_{B_2}(x)$. In this case the expressions simplify to
\al 
P_{T_S}(x) &= P_{S_1}(x) - \frac{f_2(1-f_1)}{f_1-f_2}P_{S_1} \xi_{S}(x)  \label{pts:case1} \\ 
P_{T_B}(x) &= P_{B}(x) \left(1+ \frac{f_1 f_2 \xi_S^0}{f_1-f_2(1+\xi_S^0)} \right) + P_{S_1}(x) \frac{f_1 f_2 \left(\xi_S(x) - \xi_S^0 \right)}{f_1-f_2(1+\xi_S^0)} \label{ptb:case1}
\fal
We can see that in general the signal reconstruction is better than the background reconstruction, as the expression involves only the underlying signal distribution, whereas the background topic involves both signal and background distributions in addition to denominators which may enhance the disagreement between $P_{T_B}(x)$ and $P_{B}(x)$.  More quantitative statements can be made, by calculating the distances $\Delta_{B},\Delta_{S_1}$ and $\delta_S$.   For instance, by computing $\Delta_S$ using Eq.~\ref{pts:case1} and expanding in $\xi_S$ we obtain
\begin{eqnarray}
	\Delta_{S} &=& d_H(P_{T_S},P_{S_1}) \simeq \left[\frac{(1-f_1)f_2}{f_1-f_2} \right] \frac{1}{2\sqrt2} \sqrt{\int\!dx\,P_{S_1}(x)\xi_S^2(x)} \nn \\
	&=& \left[\frac{(1-f_1)f_2}{f_1-f_2} \right] d_H(P_{S_1},P_{S_2}).
\label{distH}
\end{eqnarray}
That is, $\Delta_{S}$, a measure on the performance in the signal reconstruction, follows a linear dependence on the distance between the two signal distributions, which is a measure of the breaking of condition H2. We also see that the slope is simply a function of the signal fractions in the samples $f_1, f_2$, and that by increasing $f_1$, that is the amount of signal in sample $M_1$, this slope decreases.  This is expected, since the more pure are the samples on a given distribution, the better is the reconstruction of the corresponding topic.

To better visualize the above results, we have performed numerical simulations by scanning on different function and noise shapes. We have used Gaussian for S and B distributions, randomly sampling their means and deviations, and keeping only cases with mutual irreducibility, under a certain tolerance. For the noise functions $\xi_{S,B}(x)$ we sampled randomly for each bin a value between $[-0.1,0.1]$, then renormalizing the functions in order to satisfy relation (\ref{xinorm}) (and the analogous for the $p_B(x)$ distribution). Then the two mixture samples having fractions $f_1 = 0.45$, $f_2 =0.22$ were generated by summing the underlying distributions.

\begin{figure}[h]
\centering
\includegraphics[width=0.5\textwidth]{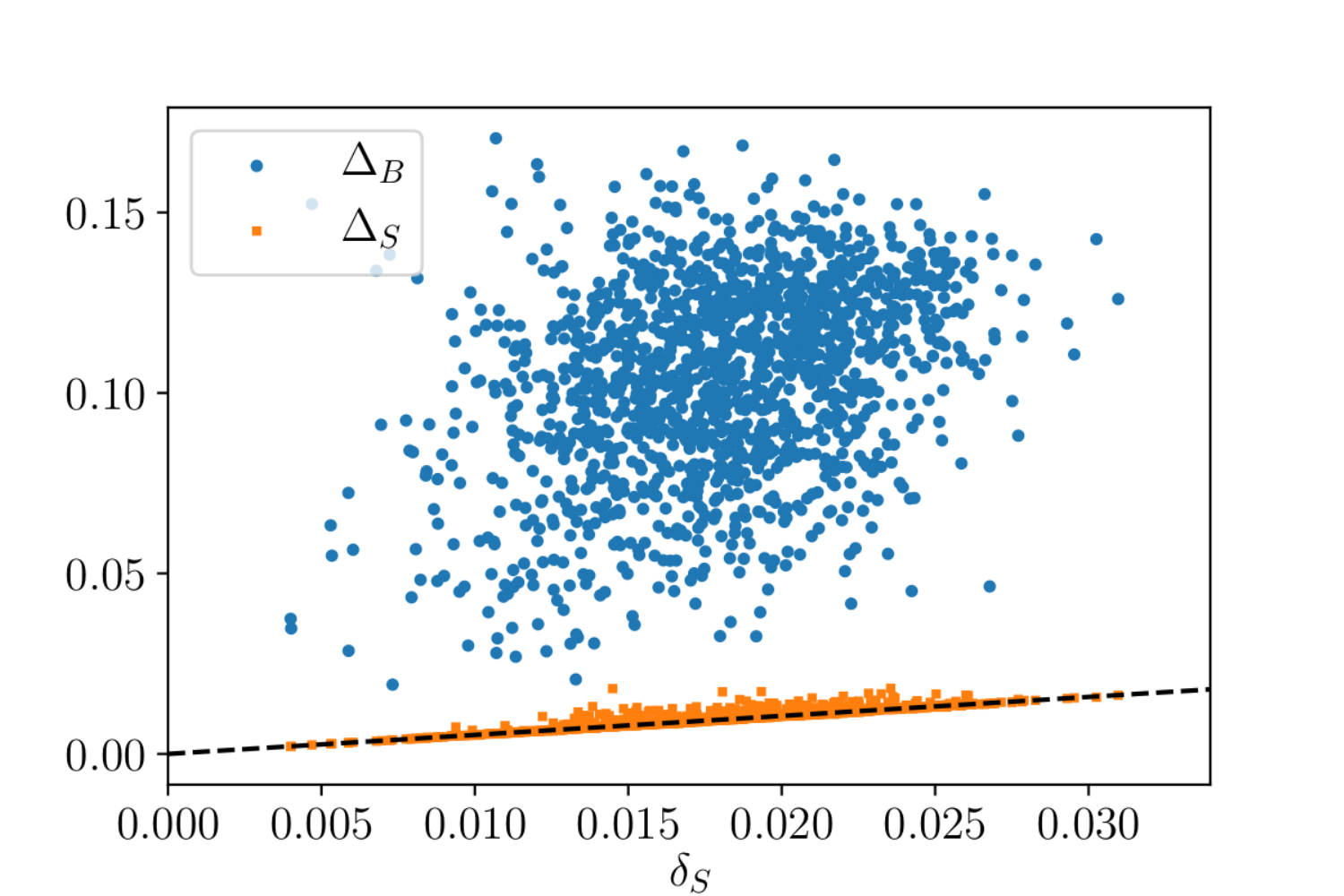}
	\caption{Testing H2 under case 1). Scatter plot of distance between the reconstructed signal and background topics and the corresponding underlying distribution as a function of distance between the two underlying signal distributions.  We show in dashed black the line given by slope $(1-f_1)f_2/(f_1-f_2)$ from Eq.~\ref{distH}.  The vertical axis corresponds to $\Delta_{S_1}$ for orange (square) points and to $\Delta_B$ for blue (circle) points.}
\label{case1}
\end{figure}

Using these simulations we have computed the above distances for each topic. Each simulated point corresponds to a pair of functions $P_{S_1}(x)$ and $P_{B_2}(x)$ mutually irreducible, and two noise functions $\xi_S$, $\xi_B$ of the same amplitude. A plot of these distances can be seen in Fig.~\ref{case1}, for case 1), which corresponds to setting $P_{B_2}(x)=P_{B_1}(x)$, that is using a single noise function $\xi_S(x)$. We see the linear dependence in signal reconstruction with the signal difference $\delta_{S}$. We have verified a similar result for other distances such as $L^1$.   This behavior is well explained by Eq.~\ref{distH}. These points can be linearly interpolated for different values of $f_1 > f_2$, to see that the slopes follow the predicted values. From Eqs.~\ref{pts:case1} and \ref{ptb:case1} one can see that the background topic is expected to be noisier.  In fact, one can see from the RHS in Eq.~\ref{distH} that for fixed $\delta_S$, the distance $\Delta_{S_1}$ is approximately constant.  Whereas the same procedure for $\Delta_{B}$ is more involved due to the other factors present in Eq.~\ref{ptb:case1} which yield stochastic noise.  We also see in Fig.~\ref{case1} that the Hellinger distance $\Delta_B$ is an order of magnitude larger than $\delta_S$, showing that the background reconstruction is sensitive to the distance between underlying signal distributions $\delta_S$.  We find less enhancement if the $L^1$ distance is used.

\begin{figure}[h]
\centering
\includegraphics[width=0.8\textwidth]{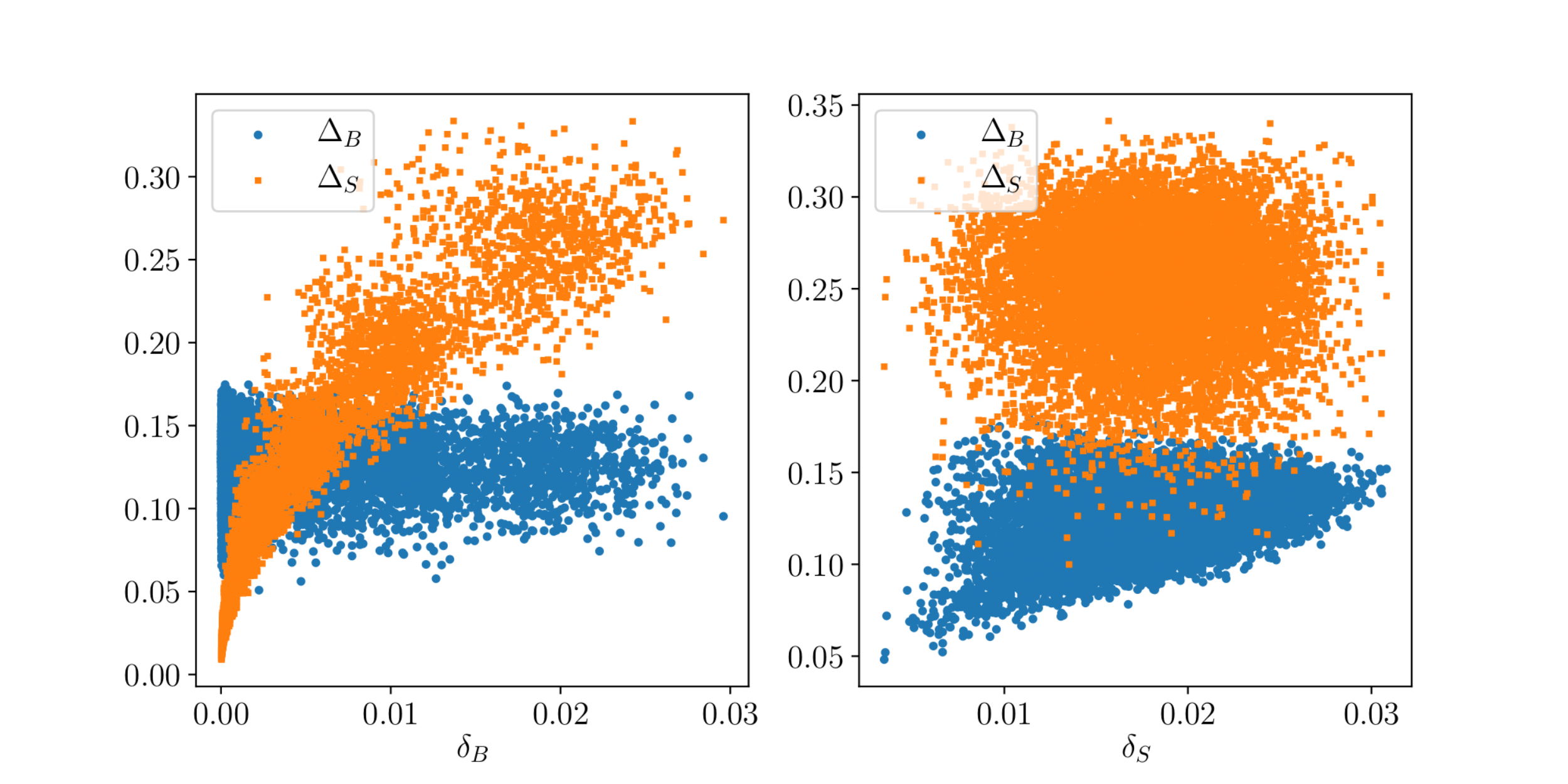}
	\caption{Testing H2 under case 2. Scatter plot for $\Delta_B$ and $\Delta_S$ as functions of $\delta_B$ and $\delta_S$, for fractions $f_1=0.45$, $f_2 = 0.22$.  In the left panel we have extended the simulations to also include values of $\delta_B$ well below $\sim 0.01$, in order to make the connection with Fig.~\ref{case1} in the limit $\delta_b \to 0$.  Vertical axis is as in Fig.~\ref{case1}.}
\label{case2}
\end{figure}

For Case 2), with different signal and background distributions in each sample $M_i$, reaching a closed form for the distances $\Delta_B$ and $\Delta_S$ is considerably more involved. We have computed numerically these distances using the same simulation scheme as in the previous case and obtained the plots in Fig.~\ref{case2}.  From using specific forms of $\xi(x)$ noise functions one can infer that the topic with more fraction in the sample is the one that would have a better reconstruction.  We have verified this statement by simulating many cases as in Fig.~\ref{case2} with different fractions.

\subsubsection*{Summary}
From the above results we see that there is an interplay between the distances of the underlying distributions and the fractions of purity of each topic in the samples to determine which topic is better reconstructed and to what extent.  For instance, in Fig.~\ref{case1} we see that having a non-zero distance between the signal distributions, $\delta_s\neq 0$, dominates and the background is less well reconstructed.  This statement is still valid for specific values of the fractions $f_i$ that increase the slope of the orange points, reaching the limit in which both topics are equally reconstructed.  On the other hand, in Fig.~\ref{case2}, we see that for $\delta_S \sim \delta_B\neq0$, the value of the fractions $f_i$ is the one that determines which topic is better reconstructed.

As a summary of the above paragraphs, we can extract some useful statements regarding the validity of the demixer when the working conditions do not fully satisfy H1 and H2.
\begin{itemize}
\item When mutual irreducibility is not guaranteed, if one of the topics has an anchor bin, then this topic is better reconstructed by the algorithm. (Case $b0$).
\item If the underlying distributions for one topic are different in the two samples $M_i$, and the other topic has equal underlying distributions, then the former is the one with better reconstruction. (Case $a1$.)
\item If both topics have different distributions over both samples, then topic reconstruction is mainly ruled by sample purities. If samples are mostly background, then background reconstruction will be better than signal reconstruction, and vice-versa. (Case $a2$.)
\end{itemize}

\section{Demixer algorithm in four-top at the LHC}
\label{section:3}

The four-top final state at the LHC is a very busy channel with little chances to be correctly reconstructed.  In addition, this signal and its background suffer important MC uncertainties.  It is therefore an attractive channel in which to apply the Demixer algorithm, and where we can expect to find many of the difficulties discussed in Section \ref{section:2}.

To choose the decaying channel in which to apply the Demixer algorithm, we note that the algorithm requires a non-negligible fraction of signal in the samples. Considering the branching ratios and background processes, we find it suitable to focus on the same-sign dilepton and tri-lepton channels where the main backgrounds are $t\overline{t}W$ and $t\overline{t}H$ with the $W$ decaying leptonically and the $H$ decaying semi-leptonically or leptonically through $WW^*$. This multilepton channel provides the highest signal over background ratio, provided we take the appropriate cuts beforehand by following Ref.~\cite{Sirunyan:2019wxt}. We have performed event MC simulations of $pp\to t \bar t t \bar t$ up to detector level using MadGraph~\cite{Alwall:2014hca} for matrix-level process, Pythia~\cite{Sjostrand:2014zea} for showering and hadronization and Delphes\cite{deFavereau:2013fsa} for detector simulation, following the same basic cuts as in Ref.~\cite{Sirunyan:2019wxt}, but at leading order and with only up to one extra parton.  We have set the mixture fractions to agree with the event yields reported in Ref.~\cite{Sirunyan:2019wxt}.

In order to have a good demixer we need to find two sets of observables as uncorrelated as possible, one to define $M_1$ and $M_2$ and the other one to play the demixing variable $x$, as defined in Section \ref{section:2}.  The uncorrelation is what provides the approximately same underlying distributions for signal and background in both samples, namely H2. A naive set of observables from which to choose is $N_b$, $N_j$ and the $p_{T}$, energy and $\Delta R$ from all the reconstructed objects in the event. In order to do a simple demixing model, we choose only one variable to define $M_1$ and $M_2$ and another one to demix. Using a multidimensional demixing with a Neural Network as in CWoLA~\cite{Metodiev:2017vrx} does not show at this extent considerable improvements.  This is discussed in Appendix~\ref{appendix}. 
\\

On physical grounds, the most direct distinction between signal and background would be the number of b-tags ($N_b$).  Observe that the SS multilepton channel has the special feature that none of the backgrounds has more than 2 b-jets.   Meanwhile, and as showed in Ref.~\cite{Alvarez:2016nrz}, the number of reconstructed jets ($N_j$) is also a good discriminator between signal and background. We consider these two variables to perform the $M_i$ definition and the demixing, leaving $p_{T}$, energy and $\Delta R$ for setting cuts to accept or reject reconstructed objects such as leptons, jets or b-jets.  Considering H2, we require the two background processes $ttW$ and $ttH$ contained in both $M_i$ to yield approximately the same underlying distribution on each sample.  Since these backgrounds have different number of jets at parton level, it is suitable at this level to avoid dividing the samples using $N_j$ because the relative proportions of these backgrounds suffer a non-negligible change in each sample, yielding different underlying distributions for the background. In addition, $N_b$ is a variable with very few discrete values to use as a demixing observable.  Therefore, we consider dividing the samples as
\begin{eqnarray}
	M_1: & & \mbox{Events with 3 b-tags,} \nn \\
	M_2: & & \mbox{Events with 2 b-tags.} \nn
\end{eqnarray}
Whereas $N_j$ is to be used as the variable on which the demixing algorithm is performed.  

\begin{figure}[h!]
\begin{center}
\subfloat[]{\includegraphics[width=0.45\textwidth]{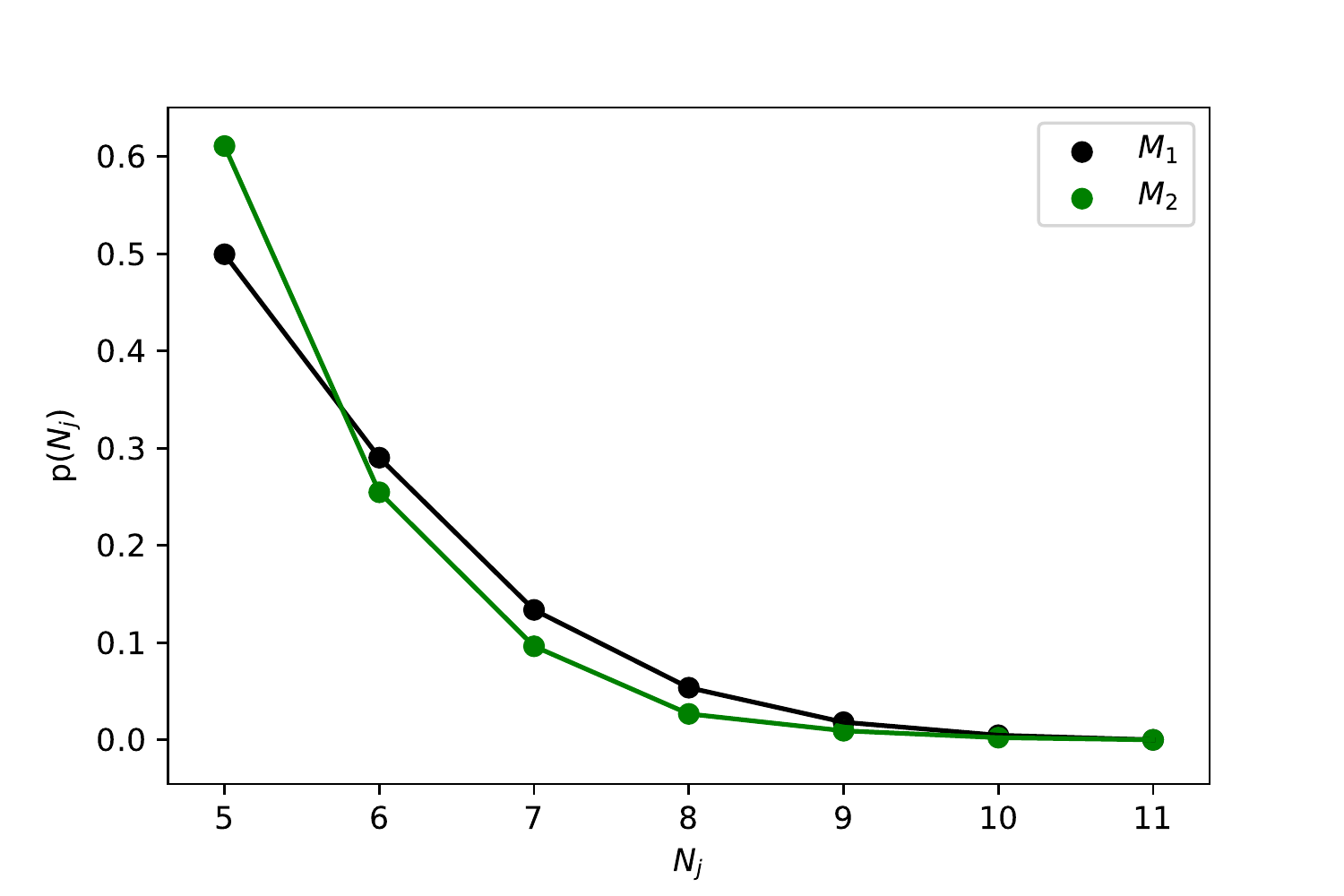}\label{fig:m1m22lep}}\hspace{3mm}
\subfloat[]{\includegraphics[width=0.45\textwidth]{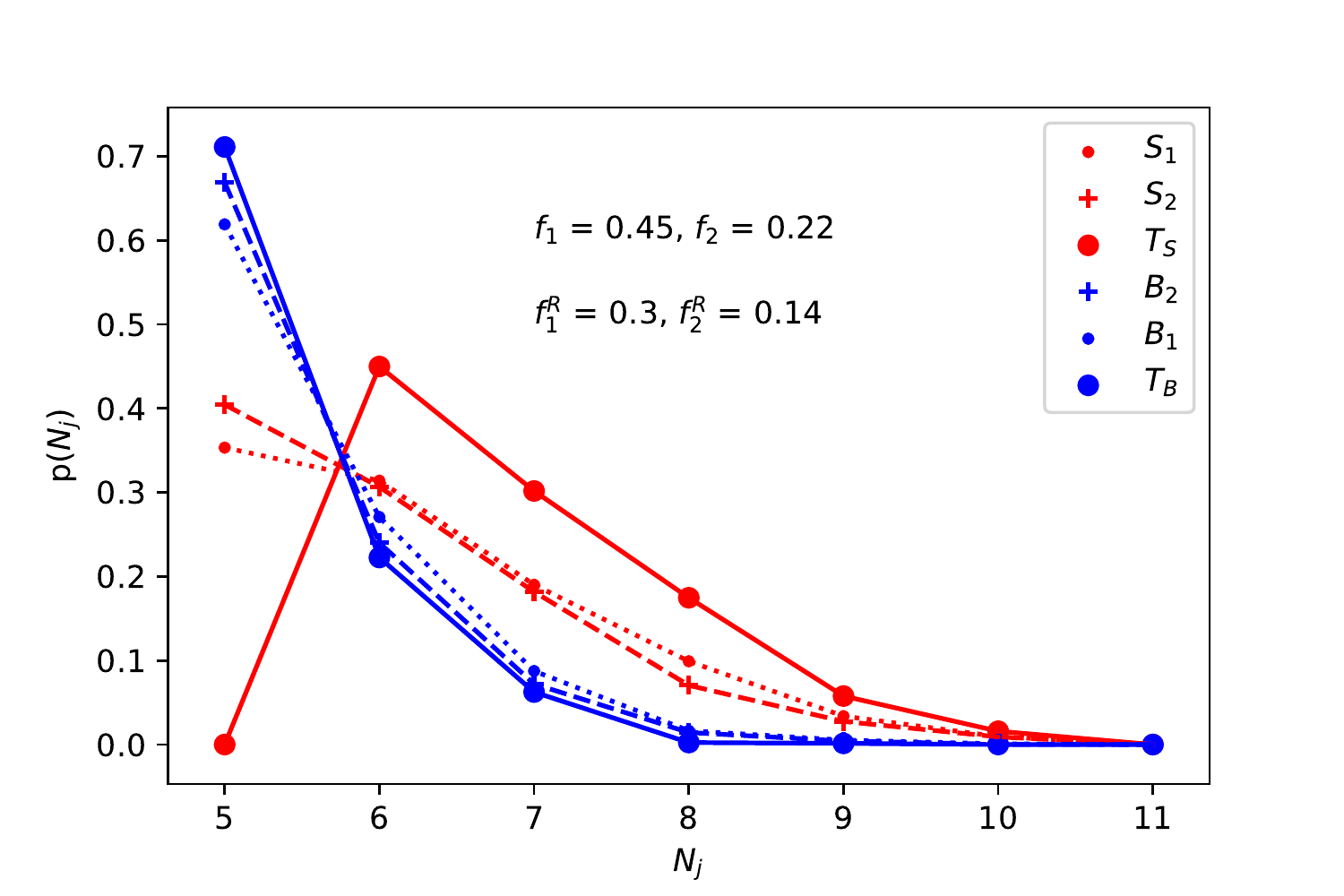}\label{fig:m1m2demixed2lep}} \\
\caption{a) $M_1$ ($N_b = 3$ ) and $M_2$ ($N_b = 2$) distribution over $N_j$. b) Reconstructed underlying distributions using the Demixer algorithm.  Solid lines represent the reconstructed topics after the demixing algorithm is performed on the left panel samples.  We also show in dotted and dashed the truth-level underlying distributions in each mixture sample.}
\label{fig-demix}
\end{center}
\end{figure}

Using the above $M_i$ definition we can construct the two mixture samples from the simulated events.  We show in Fig.~\ref{fig:m1m22lep} the distribution of $M_i$ in $N_j$.  As expected, the $M_1$ mixture --which has the largest fraction of signal-- is shifted towards large $N_j$ with respect to the $M_2$ mixture.  We have computed that for full LHC Run 3 the two distributions would be distinguishable at a $\sim3\sigma$ level at the $N_j=7$ bin.

We can perform the demixing algorithm as described in previous section assuming, in a first step, that it is an ideal case.  Since we can expect from general grounds that the background distribution goes to zero for large $N_j$, we can predict a good reconstruction for the background.  We plot the reconstructed underlying probability densities as well as the truth-level topic distributions in Fig.~\ref{fig:m1m2demixed2lep}.  In fact, the background underlying distributions are approximated properly by the background reconstructed topic through the algorithm.  As studied in Section \ref{section:2}, this is because the background has a proper anchor bin at large $N_j$, it has similar distributions in $M_1$ and $M_2$, and also because it has larger fraction than signal in both samples.  On the other hand, the reconstructed signal topic does not match the underlying signal distributions because of the lack of an anchor bin.   As discussed below, obtaining a trustable background distribution in the signal region from data provides a new way of tuning the Monte Carlo event generator in the signal region.

Further analysis of the demixing algorithm indicates that the reconstructed fractions misestimate their true values by about a $\sim 30\%$.  This result is obtained through a MC-independent algorithm and the shift is within the order of magnitude of the usual MC normalizations performed in four-top signal and background predictions.   Moreover, this result is the product of applying the demixing assuming H1 holds, which we know is not true from theoretical grounds: signal does have events in all the $N_j$ bins.  As a matter of fact, we can see in Fig.~\ref{fig:m1m2demixed2lep}, that it yields a background-subtracted signal distribution which assigns a zero probability to $N_j = 5$.  Therefore, we can still improve this result, but at the price of including Monte Carlo input.

\begin{figure}[t!]
\begin{center}
\subfloat[]{\includegraphics[width=0.45\textwidth]{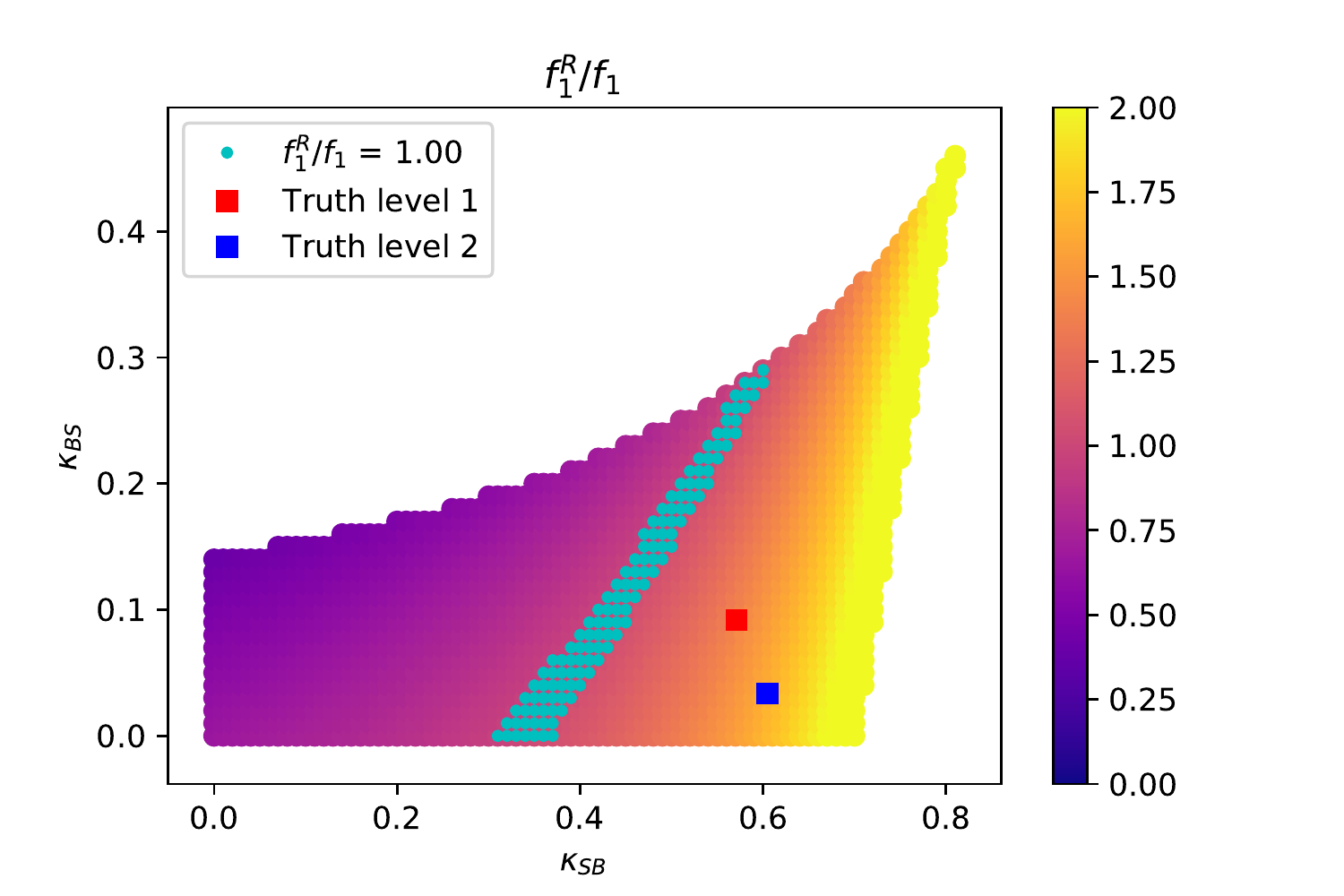}\label{fig:kplane2lep}}\hspace{3mm}
\subfloat[]{\includegraphics[width=0.45\textwidth]{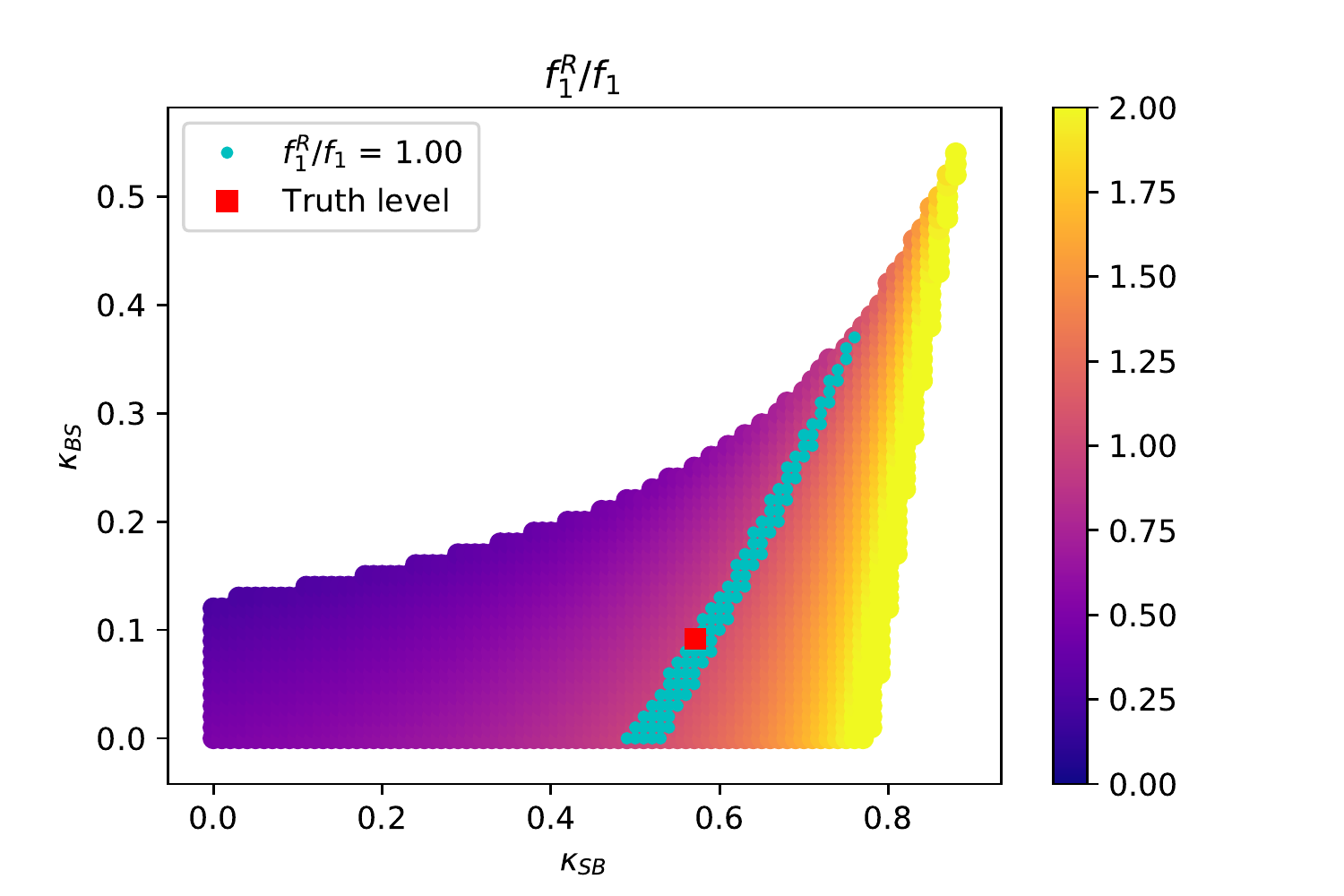}\label{fig:kplane2lep_sameunder}}\hspace{3mm}
	\caption{Reconstructed $f_1^R$ modified by $\kappa_{SB}$, $\kappa_{BS}$ in terms of the true fractions for a) different underlying distributions and b) same underlying distributions. The truth-level ($\kappa_{SB}$,$\kappa_{BS}$) pair are estimated using the underlying distributions in $M_1$ and $M_2$.}
\label{fig:kplane}
\end{center}
\end{figure}

\begin{figure}[ht!]
\begin{center}
\subfloat[]{\includegraphics[width=0.45\textwidth]{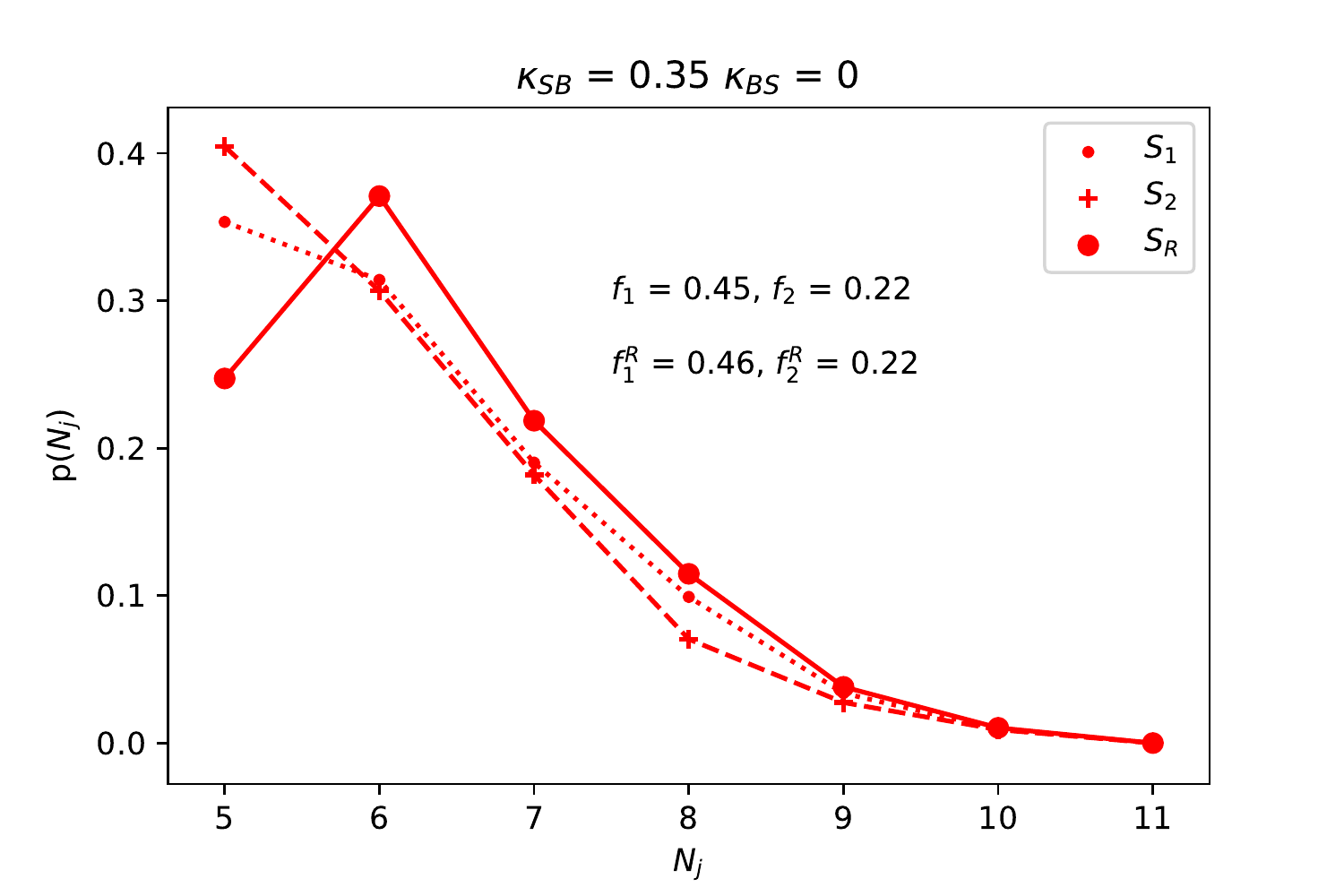}}\hspace{3mm}
\subfloat[]{\includegraphics[width=0.45\textwidth]{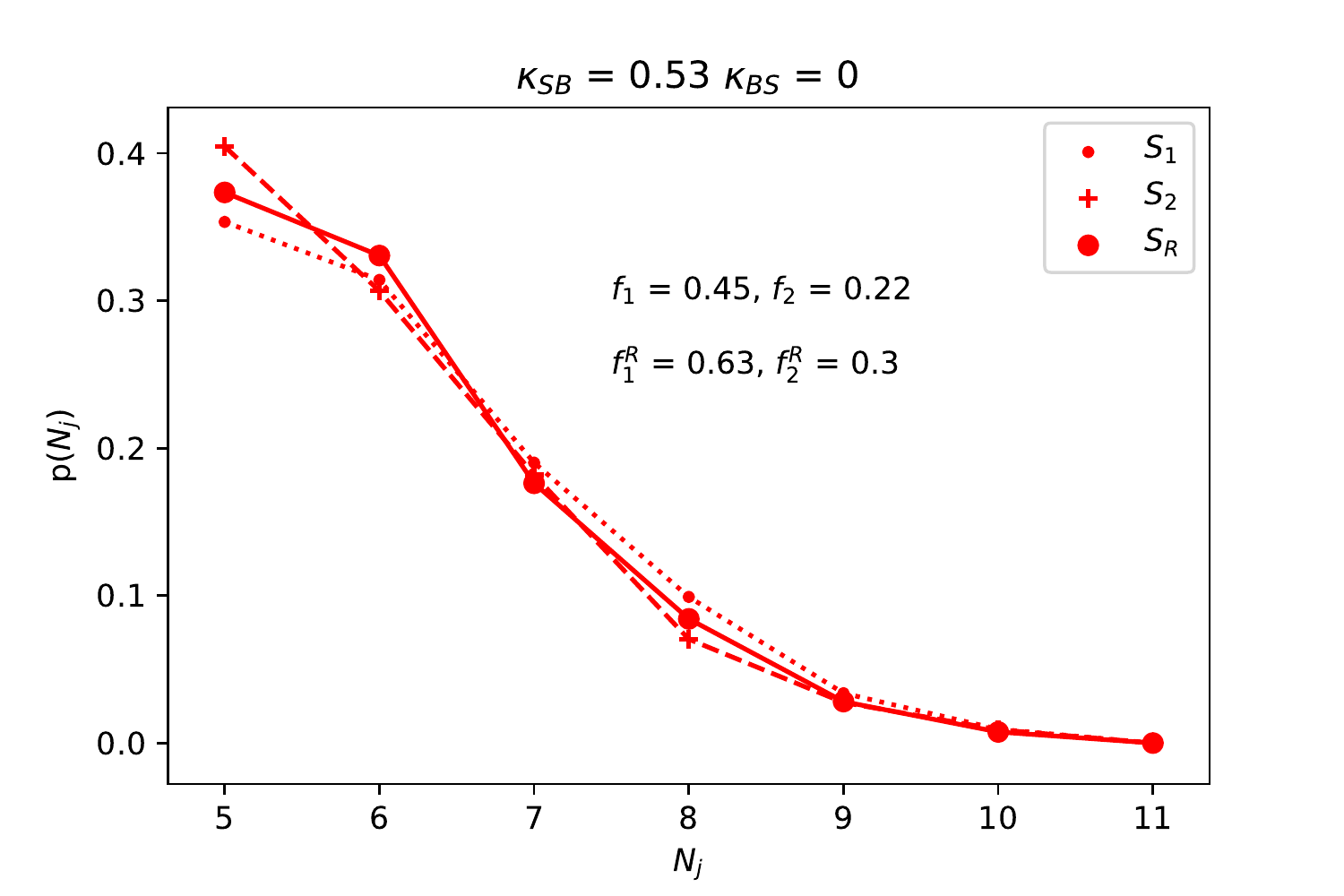}}
	\caption{Manually increasing of $\kappa_{SB}$ while leaving $\kappa_{BS}=0$, yields at some point the correct fractions but disagreement with the distributions (left panel), whereas further increasing $\kappa_{SB}$ yields correct distributions, but incorrect fractions (right panel). The truth-level distributions are $S_1$ and $S_2$ whereas $S_R$ is the distribution obtained using Eq.~\ref{substracted}. This is an explicit and graphical demonstration on the sensitivity of the algorithm to H2.}
\label{fig:2lep_tuned}
\end{center}
\end{figure}

As a second step, we address the demixing algorithm using the workaround for the absence of anchor bins, the $\kappa$ factors defined in Section \ref{section:2}. \change{These $\kappa$ factors could be understood as hyperparameters in the algorithm, since a prior knowledge or constraint on their possible values provides a better adjustment in the estimated fractions. We can study the performance in the reconstructed fractions using the hyperparameter plane ($\kappa_{SB}$, $\kappa_{BS}$).}  In Fig.~\ref{fig:kplane} we scan on these $\kappa$ for the real case of different underlying distributions in both samples (Fig.~\ref{fig:kplane2lep}) and for the adjusted case of equal underlying distributions in both samples (Fig.~\ref{fig:kplane2lep_sameunder}).  \change{In both cases we see that using the prior theoretical knowledge that $\kappa_{SB}>0$, and tuning  manually $\kappa_{SB}$ to larger values while leaving $\kappa_{BS}=0$, pushes $f_1^R / f_1$ to one.}  Moreover, we can see that when the underlying distributions are not equal in both samples (Fig.~\ref{fig:kplane2lep}), the $\kappa$'s that correctly reconstruct $f_1^R / f_1$ do not coincide with the solution corresponding to the correct underlying distributions.  It is interesting to see in Fig.~\ref{fig:2lep_tuned} how by manually tuning $\kappa_{SB}$ to larger values we reach a solution in which the fractions are correctly reconstructed, but not the distributions, and then vice-versa.  This behavior is still obtained if one also varies $\kappa_{BS}$, since the reason behind this disagreement is that H2 is not fulfilled.  

\begin{figure}[ht!]
\begin{center}
\includegraphics[width=0.45\textwidth]{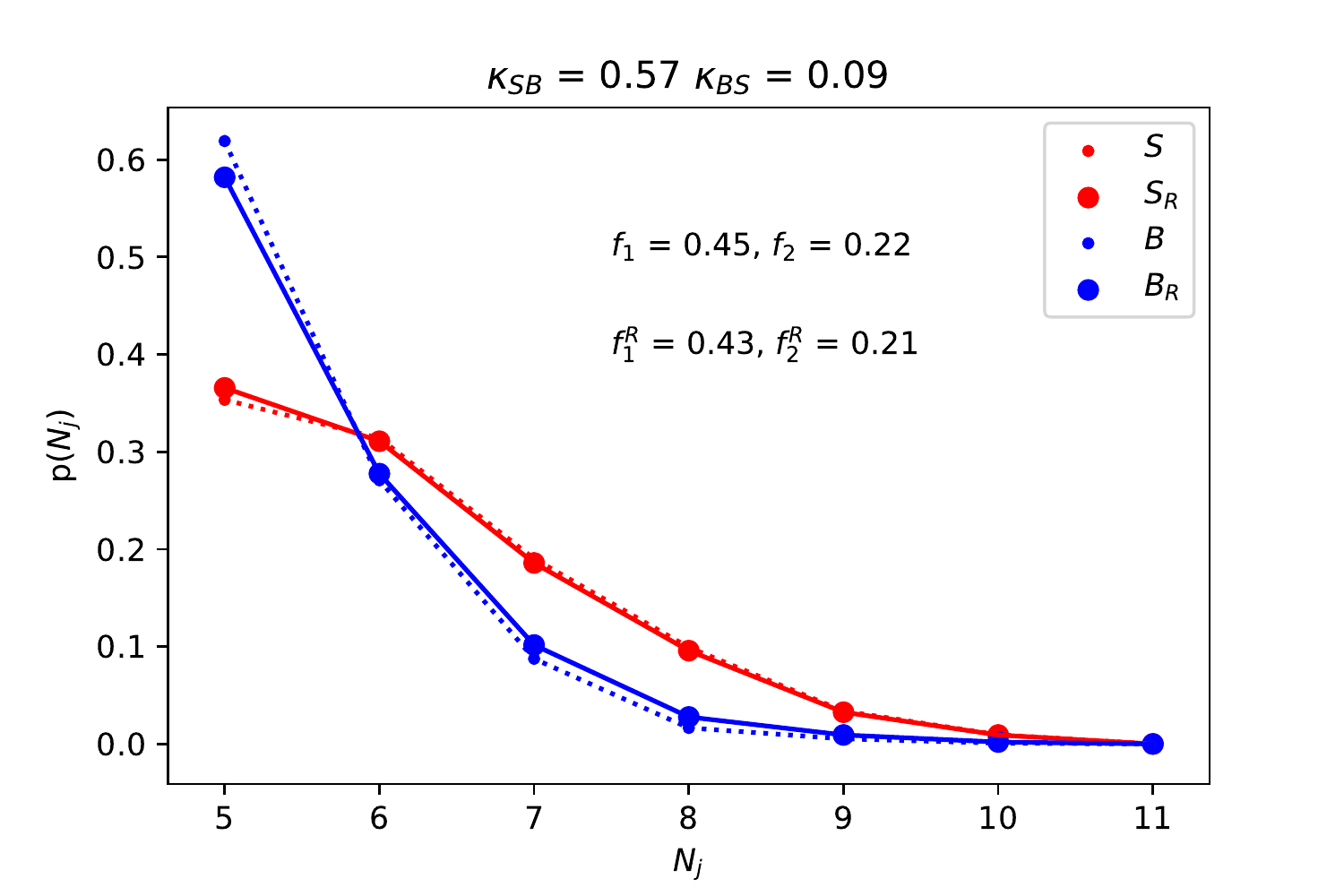}
	\caption{Corrected distribution and fractions for the truth-level ($\kappa_{SB}$, $\kappa_{BS}$) when forcing the underlying distributions to be the same. Here $S_R$ and $B_R$ stand for the distributions obtained using Eq.~\ref{substracted}.}
\label{fig:2lep_perfect}
\end{center}
\end{figure}

If, on the other hand, we satisfy H2 by forcing to have same underlying distributions (Fig.~\ref{fig:kplane2lep_sameunder}), we can \change{tune the hyperparameters $\kappa_{SB,BS}$ to correctly reconstruct the fractions and the distributions.  For the sake of completeness, we show in Fig.~\ref{fig:2lep_perfect} the output for the demixing algorithm in this case. }   The plots in Fig.~\ref{fig:kplane} show the sensitivity of the algorithm to H2, as noticed in previous section.  We also see in this figure that the correct solution is more sensitive to $\kappa_{SB}$ than to $\kappa_{BS}$.   This is expected, since the background does have an (approximate) anchor bin, and thus $\kappa_{BS}$ is expected to be close to zero.  

In the following section we complete this discussion with possible methods to combine the demixing algorithm with MC simulations to reduce the impact of MC tuning in the extraction of physical quantities from observations.

\section{Discussion}
\label{section:4}

Along previous sections we studied how the Demixer algorithm works in recovering signal and background distributions from mixture samples, and how to apply these tools to the four-top process. In this section we discuss the strengths and some of the shortcomings of this implementation, along with possible improvements.  We first discuss possible goals relative to the analysis in Section \ref{section:2}, when the Topic Model hypotheses are not fully satisfied.  We then discuss possible variations to the Demixer algorithm implementation presented in Section \ref{section:3} together with its pros, cons, and features to be further explored.  We end with a discussion on how the presented Demixer algorithm could be implemented to extract physical quantities in four-top while reducing the impact of MC simulations.

In Section~\ref{section:2} we detailed the hypotheses behind the topics and fractions extraction, exploring what happens when these hypotheses are not valid. A more exhaustive study can be made, both analytically and numerically. For instance, the error propagation of the demixer algorithm can be studied for the more realistic case in which the uncertainty in each sample is taken into account. Error bars in samples $P_{M_1}(x)$ and $P_{M_2}(x)$ would be translated into error bars in $P_S(x)$, $P_B(x)$, along with reconstructed fractions $f^R_{1,2}$. Analyzing this error propagation and its dependence on the validity of conditions H1 and H2 defined in Section \ref{section:2}, would be an important step towards better understanding the algorithm and its potential.

In Section~\ref{section:3} we used the SS multilepton channel for applying the Demixer algorithm to the four-top process.  The main reasons for this choice are the relatively high $S/B$ ratio, and also that this channel has the special condition that all background processes consist of at most two b-jets at parton level.  It is worth noticing at this point that other minor backgrounds to the SS channel, such as non-prompt leptons and $t\bar t Z$, have a  behavior similar to the main backgrounds in what concerns to the number of $b$-jets and also to the anchor bin for large $N_j$.   We have also studied other channels such as mono-lepton, for which we only mention the following results.  We find that being $S/B$ below ${\cal O}(10^{-2})$ makes it challenging to correctly apply the algorithm.  Moreover, even in the hypothetical case that new cuts could increase $S/B$ we find that using $N_b$ to define the samples and $N_j$ to demix yields different underlying background distributions, which breaks H2 and therefore spoils the results.  In fact, since in this case $t\bar tb\bar b$ is among the main backgrounds, the relative contribution of this background to the background topic would change considerably between the $N_b=2$ and $N_b=3$ samples.  Despite of these difficulties in approaching this channel with the Demixer algorithm, we still find it an interesting goal, since having a less MC-dependent prediction of backgrounds such as $t\bar t b \bar b$ is an important avenue in four-top and in heavy flavor physics.

Along the article we have used b-tagging information through $N_b$ to define the samples $M_i$, and $N_j$ to demix.  An inversion of the roles of these variables would be an interesting study.  Of course, in this case one could not demix using $N_b$ since it can take very few discrete values.    In fact, to explore an inversion of roles one should define the samples as --for instance-- $M_1$: $N_j>7$ (signal enhanced) and $M_2$: $5 \leq N_j \leq 7$ (signal suppressed), and define a continuous variable considering the probability of having b-tags within the jets.  This variable could be, for instance, the sum of the MV2c20 \cite{mv2} variable\footnote{This and/or similar variables, and its subsequent development, are the optimal variables on which the probability density of a light-jet and a b-jet are best discriminated. These are the variables used to define the b-tagging working points.} of the four leading jets.   Such a variable would permit to perform a demixing on this variable, and bring up a different prediction for four-top using the Demixer algorithm. 

\begin{figure}[h!]
\begin{center}
\subfloat[]{\includegraphics[width=0.45\textwidth]{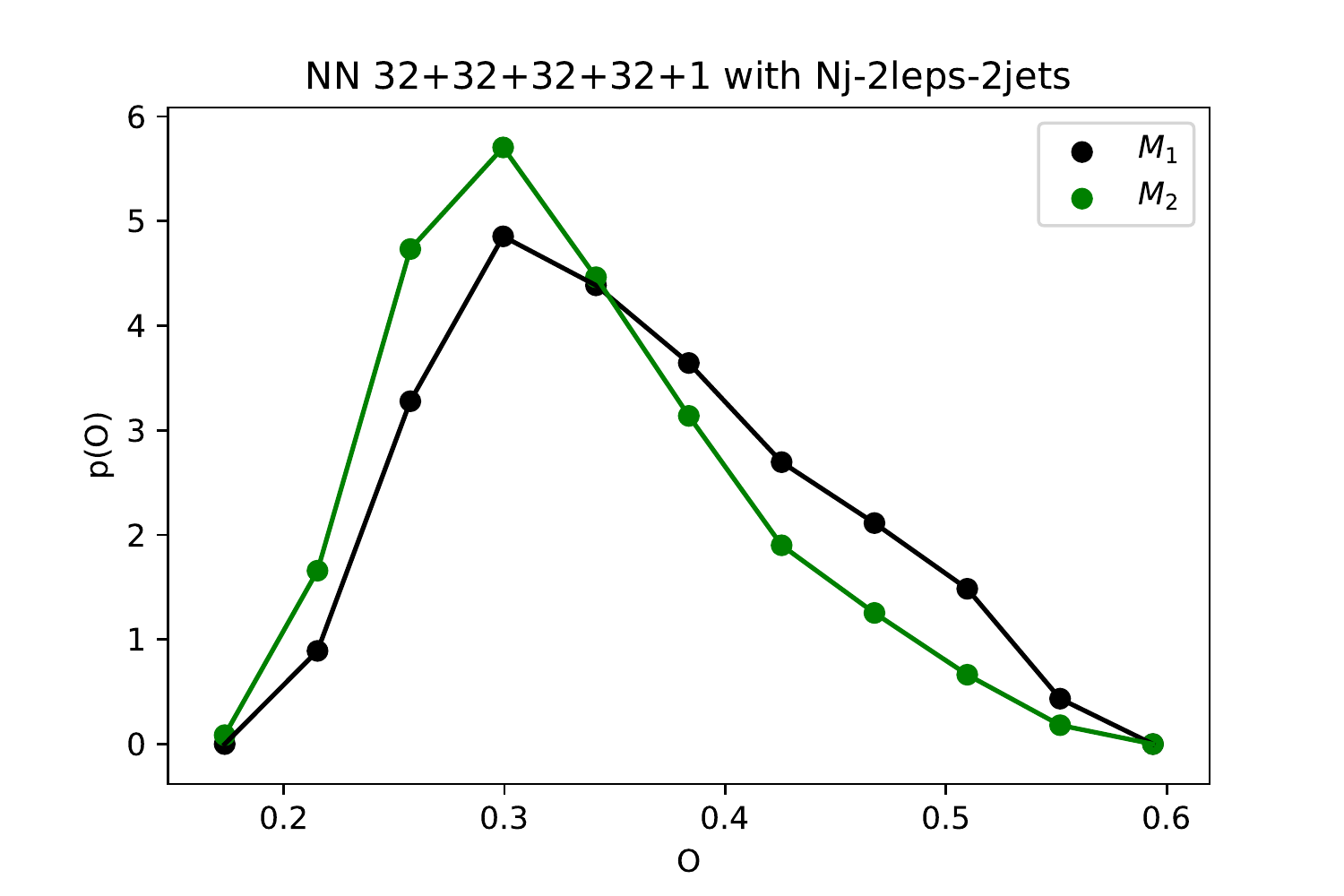}}\hspace{3mm}
\subfloat[]{\includegraphics[width=0.45\textwidth]{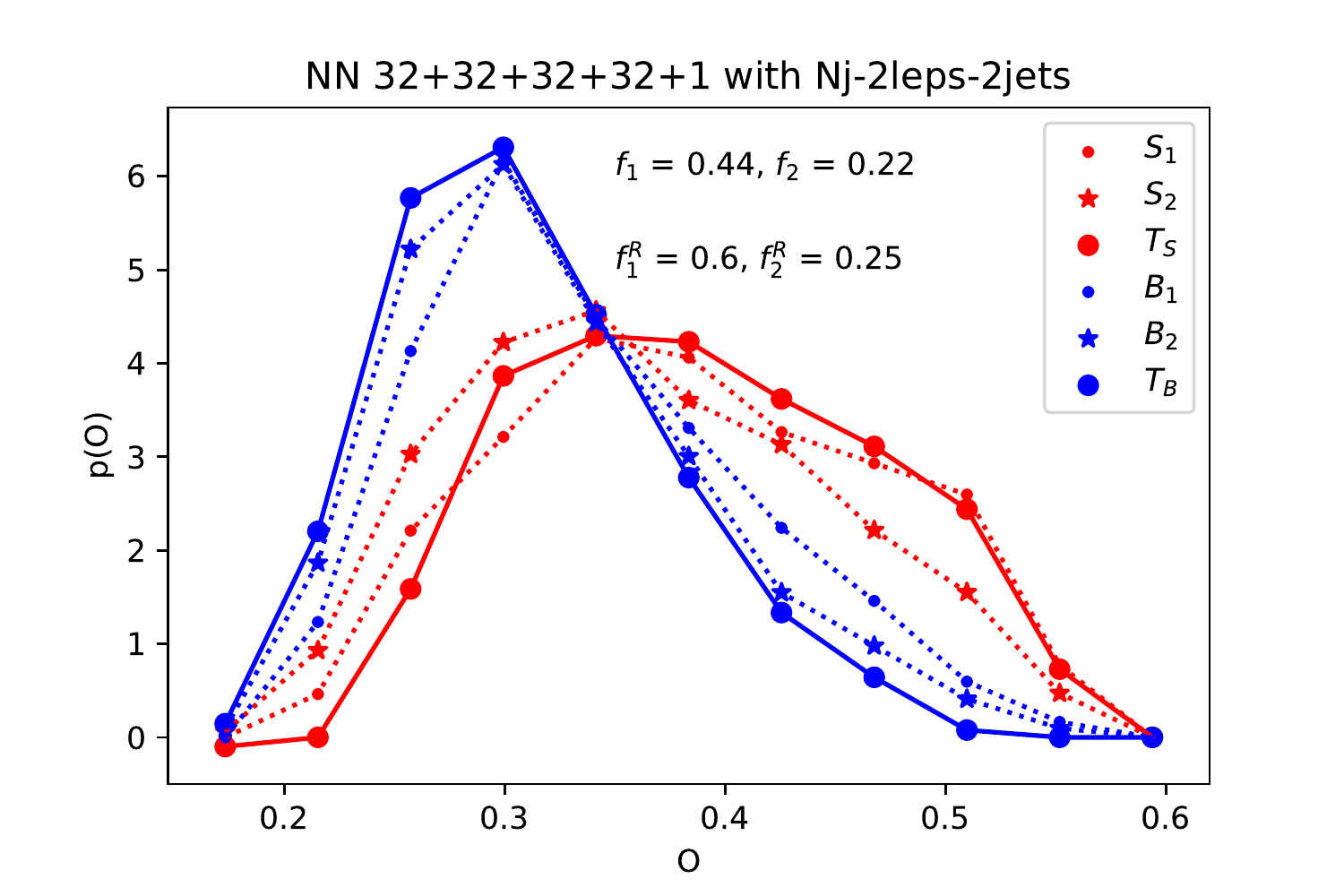}} \\\hspace{3mm}
	\caption{Demixer algorithm through a Neural Network (32+32+32+32+1, see Appendix \ref{appendix}) trained to distinguish $M_1$ from $M_2$ using as input $N_j$ and the four-momentum information of the two leading leptons and two leading jets.  Left: $M_1$ and $M_2$ distribution on the Neural Network output $o$. Right: topics reconstruction and original underlying distributions in the same Neural Network output.  These results should be compared to Fig.~\ref{fig-demix}, which is the version of this figure using $N_j$ instead of the Neural Network output.}
\label{fig:ml}
\end{center}
\end{figure}

We have also implemented Machine Learning techniques to explore the ability of a Neural Network to best discriminate $M_1$ from $M_2$, and therefore signal from background.  The bottleneck in this direction is to choose the parameters for the Neural Network whose distribution remains the more similar possible between $M_1$ and $M_2$, otherwise the breaking of H2 spoils any improvement that could be done with the Neural Network.  We have discussed different architectures and choice of parameters in Appendix \ref{appendix}, where we show the demixing result for each case.  In Fig.~\ref{fig:ml} we show the result for teaching a Neural Network to distinguish $M_1$ from $M_2$ and then using its output as the demixing variable.
Losing physical interpretation on the demixing variable to assure same underlying distributions, yields potential issues in guaranteeing H2, which is translated into a more challenging topic reconstruction.  More details on the procedure and the architectures can be found in Appendix \ref{appendix}.  We find moderate improvements compared to demixing in $N_j$ alone, meaning that $N_j$ is the main discriminator between signal and background at this level.  \change{It is also worth noting that the Demixer algorithm is a data-driven technique that does not require training and, using eight i7 CPU cores, it performs with times of order $\mathcal{O}(10^{-3})$ seconds while the CWoLA algorithm requires times of order $\mathcal{O}(10^{2})$ seconds for a simple architecture and a relatively small dataset as implemented in Appendix~\ref{appendix}.  One of the reasons for this difference is the small number of bins over which we do the demixing.} Nevertheless, we consider that more work in this direction, including a continuous variable for the b-tagging information, could provide still further improvements to the main result in this work.  

We have seen along the work the sensitivity of the algorithm to hypothesis H2.  Since we are including different backgrounds altogether in the background topic, this makes it easier to break H2 and having different underlying distribution in each sample.  This is in particular a crucial failure for the mono-lepton channel briefly discussed above.  To tackle this issue a generalization of the demixer into --for instance-- three samples, now to distinguish between the backgrounds themselves, would be an interesting work to address. 

We end this section with a brief discussion on how the obtained results using the Demixer algorithm on four-top could be used to reduce the impact of MC simulations and its tuning in the extraction of an absolute physical quantity such as the four-top production cross-section, $\sigma(t\bar t t \bar t)$.  We consider one specific possibility, however there are different options which could be explored.   As discussed in the introduction, we consider that it is not possible to avoid a MC dependence to measure a quantity such as $\sigma(t\bar t t \bar t)$, which can be compared to theoretical matrix level calculations.  However, reducing the impact of these predictions and tunings is a crucial task in the four-top final state which has many complex ingredients such as ISR/FSR, hadronization, jet reconstruction and isolation, among others, which still need to be further understood.

The Demixer algorithm applied to four-top as implemented in previous section yields a reasonably reliable distribution for the background topic, since the background fairly satisfies having an anchor bin and has more purity in both $M_i$ samples.  This is an important result because it provides a MC-independent prediction for the background in a region where signal is expected.   Therefore, one would be implementing a data-driven MC tuning for the background in the signal region. That is, to tune the MC to reproduce the background shape in $N_j$ extracted from the Demixer algorithm.   In contrast to usual MC tuning in control regions, where there is no signal, this kind of tuning has the advantage that it does not require an extrapolation to different regions in parameter space.  Once this is performed, one could use this MC to predict the $\kappa_{SB}$ and from this number extract the shape of the signal distribution in $N_j$ from the Demixer algorithm.  A comparison of the signal shape predicted by the tuned MC and the shape predicted by the Demixer once the $\kappa_{SB}$ has been determined, would be a measure of the success of the method.  In such a case, one could extract the fractions of signal and background in $M_i$ and rely on this tuned MC to extract the $t\bar t t \bar t$ production cross-section.    

\section{Conclusions}
\label{section:5}
We have studied the Demixer algorithm as well as some of its limitations, and applied it to the four-top signal and its background at the LHC, where many of the required hypotheses are not fully satisfied.

In real case scenarios the requirements for applying the Demixer algorithm are usually not exactly satisfied.  We have analyzed how the outcome of this algorithm is affected when some of the required assumptions break down.  We find that, to some extent, the method is still useful.  We have shown explicitly how the topics reconstruction and fraction estimations may shift because of not having equal underlying distributions in the initial mixture samples and/or if there is not mutual irreducibility. The study is not exhaustive, and we conclude that more work in this direction is needed to better understand the Demixer algorithm and its scope beyond ideal cases.

We have implemented the Demixer algorithm for the $t\bar t t \bar t$ final state and its main backgrounds.  We have used the Same-Sign multilepton channel which assures a reasonable $S/B$ and has the special feature that all backgrounds have no more than two b-jets in their final state.   The implementation has been made using simulated events up to detector level.  We have defined the two mixture samples using the number of b-tags, $N_b=2$ and $N_b=3$, while we have used $N_j$ to demix the samples.    We have also essayed to demix using Machine Learning through a Neural Network which best discriminates the mixed samples, as in CWoLA. We find that Neural Network results can be slightly better, but at the price of considerably hiding the clarity of just using $N_j$. Neglecting statistic and systematic uncertainties we have shown that the reconstruction of signal and background topics is as predicted by the Demixer algorithm beyond the ideal case:  {\it i)} A lack of anchor bin in signal yields that only background is correctly reconstructed; {\it ii)} Signal is correctly reconstructed provided the corresponding reducibility $\kappa$-factor, which requires input beyond data-driven; and {\it iii)}  The estimation of the fractions and reconstruction of the topics yield values somewhat shifted from its true values due to slight differences in the underlying distributions.  The Demixer algorithm in the described framework without using MC inputs predicts the fraction of signal in the mixture samples with a misestimation of approximately $\sim 30 \%$.

We have discussed some possible future directions and improvements regarding the results in the article, as for instance to invert the roles of $N_b$ and $N_j$ for defining the samples and demixing, replacing $N_b$ by a more continuous b-tagging variable.  We have proposed to consider the Demixer algorithm to obtain a MC-independent prediction for the four-top background, which would allow  to tune the MC parameters for the background prediction in the signal region.  This tuned MC could be used as a new tool to extract physical quantities from the signal region, as for instance the $t\bar t t \bar t$ production cross-section.  Achieving in this way the general goal of this work, which is to reduce the impact of MC in the extraction of physical quantities.

The four-top is a very challenging final state within any framework and the Demixer algorithm is not the exception.   The actual implementation of the algorithm for four-top at the LHC along the lines presented in this manuscript would require the full LHC Run 3 dataset, and still many experimental aspects and uncertainties would have to be further analyzed.  Among the main issues to be addressed in a real implementation, we can mention that a better discriminating variable than $N_j$ that could provide an improved differentiation between the mixture samples would be a crucial milestone.    We present the results in this article as an alternative step towards reducing the impact of MC generators in the extraction of physical quantities in four-top physics.

\section*{Acknowledgments}
We thank Jesse Thaler for very useful conversations during the development of this work.  E.A.~thanks to participants of Voyages Beyond the SM III Workshop for useful discussions.

\appendix

\section{Using NNs for CWoLA implementation in 4-tops}
\label{appendix}
\setcounter{equation}{0}

In Section~\ref{section:3} we detail an implementation of Demixer algorithm to demix four-top from two backgrounds in the Same-Sign lepton channel. Demixer algorithm, as detailed in Section~\ref{section:2}, is implemented in a simple and clear way, with only one observable ($N_b$) to define the orthogonal regions $M_1$ and $M_2$ and one observable ($N_j$) to demix and obtain the topics $T_S$ and $T_B$.
In principle this could be improved as detailed by CWoLA~\cite{Metodiev:2017vrx} with the use of a larger set of observables that have to be combined in some way to get an optimal classifier for $M_1$ and $M_2$ which corresponds to the optimal classifier for signal and background. However, to identify the output of CWoLA to the real signal and background distributions, one still has to consider the validity of H1 and H2 in Section \ref{section:2}

In this appendix we study the use of Neural Networks (NN) to search for a better discriminant than using only $N_j$, while we maintain the definition for $M_1$ and $M_2$ as a function of the number of $N_b$.   We implement our algorithms through the Keras package~\cite{chollet2015keras} for Python 3.  This discriminant should provide a better resolution between $M_1$ and $M_2$ while also providing anchor bins for both the reconstructed distributions.  These distributions should be identified with signal and background if the hypotheses in Section \ref{section:2} are fulfilled.  

We feed the NN with the same simulated events, and label the events according to their classification into $M_1$ or $M_2$.  We test different NN architectures and different observables of the reconstructed events. \change{We split the samples into training, validation and testing samples. The training is performed using 200 epochs provided there is no overfitting. We make use of the classes weights option provided by Keras to account for the different number of events for $M_1$ and $M_2$ reported in Ref.~\cite{Sirunyan:2019wxt} without discarding any simulated event.} From each event we extract $N_j$, the $p_{T}$ and energy of all reconstructed objects, angular distance between any pair of reconstructed objects, and total transverse energy $H_T$. In the following we use layers of neurons with {\tt ReLu} activation functions and a final neuron with a sigmoid activation function forming a feed-forward NN trained with a binary cross-entropy loss function. The notation for the chosen observables is $N_1\text{object}_1$-$N_2\text{object}_2$ which means that we use the $p_{T}$ and energy of the $p_T$-leading $N_1$ and $N_2$ objects $\text{object}_1$ and $\text{object}_2$ and the $\frac{1}{2}(N_1+N_2)(N_1+N_2-1)$ angular distances between all of them.  For instance, in ``Nj-2leps-2jets'' we refer that the NN is being fed with $N_j$ and all the information of the 2 leading leptons and 2 leading jets in form of $p_T$, energy and angular distances between them.

\begin{figure}[h!]
\begin{center}
\subfloat[]{\includegraphics[width=0.45\textwidth]{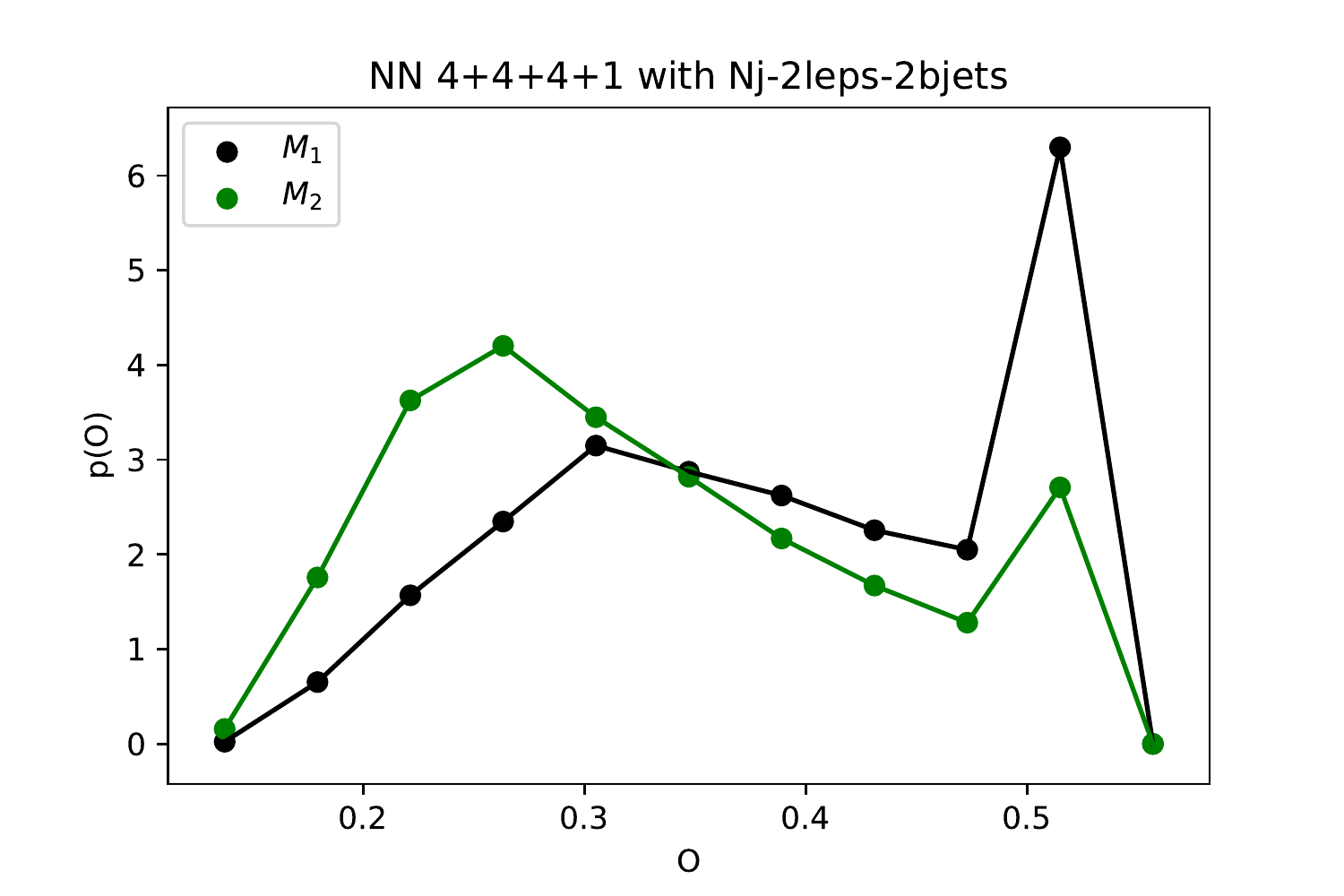}}\hspace{3mm}
\subfloat[]{\includegraphics[width=0.45\textwidth]{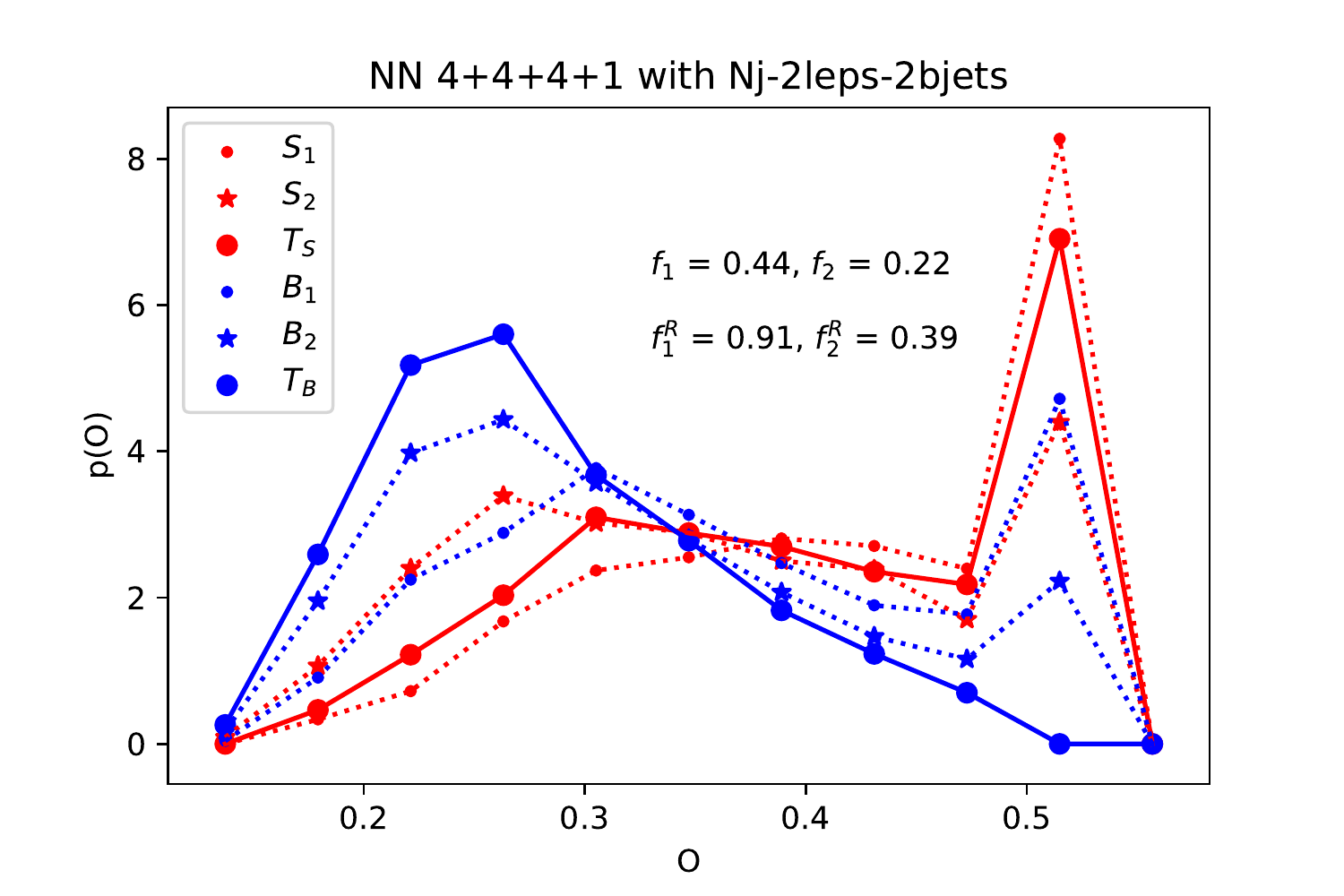}} \\\hspace{3mm}
\subfloat[]{\includegraphics[width=0.45\textwidth]{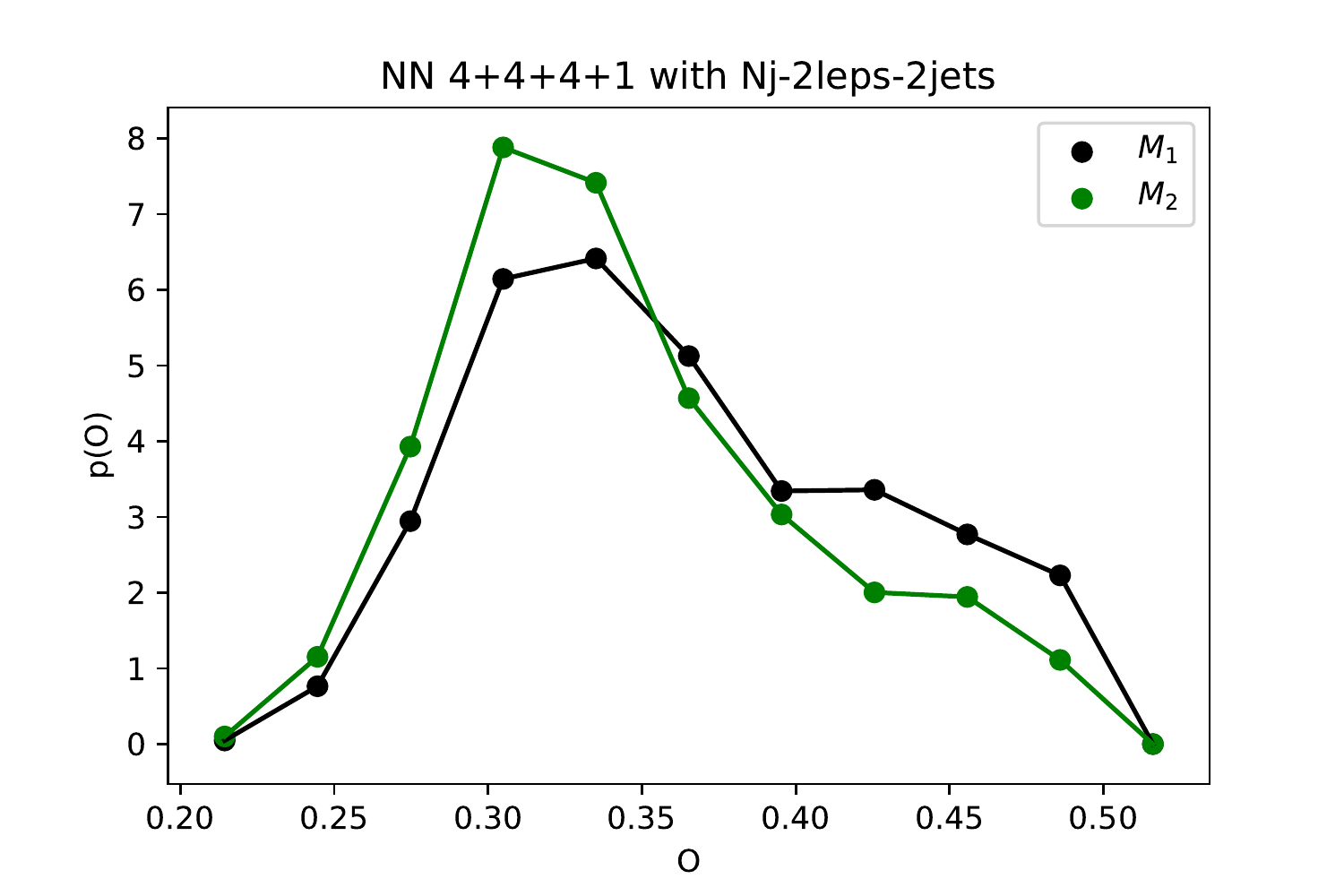}}\hspace{3mm}
\subfloat[]{\includegraphics[width=0.45\textwidth]{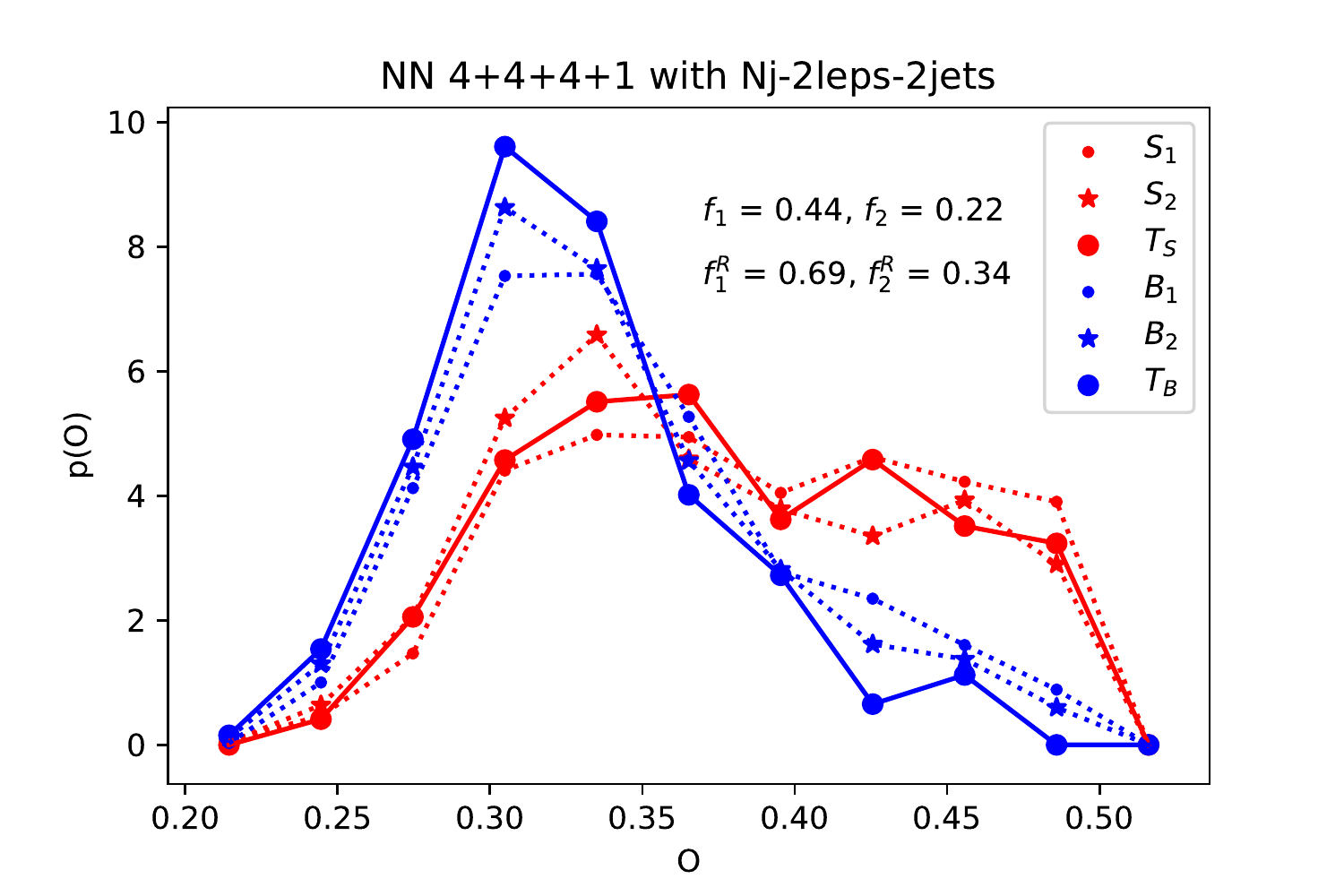}} \\\hspace{3mm}
\subfloat[]{\includegraphics[width=0.45\textwidth]{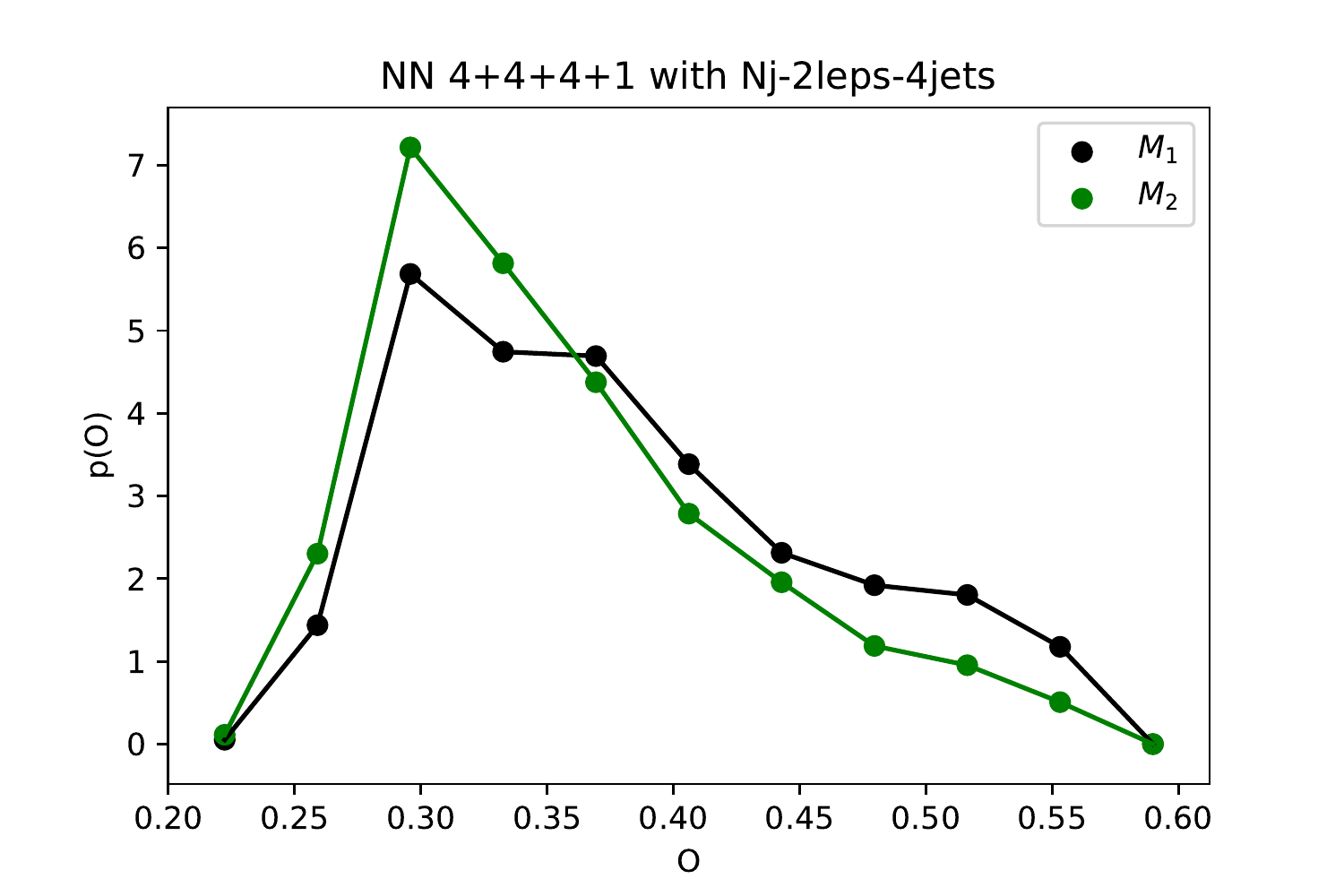}}\hspace{3mm}
\subfloat[]{\includegraphics[width=0.45\textwidth]{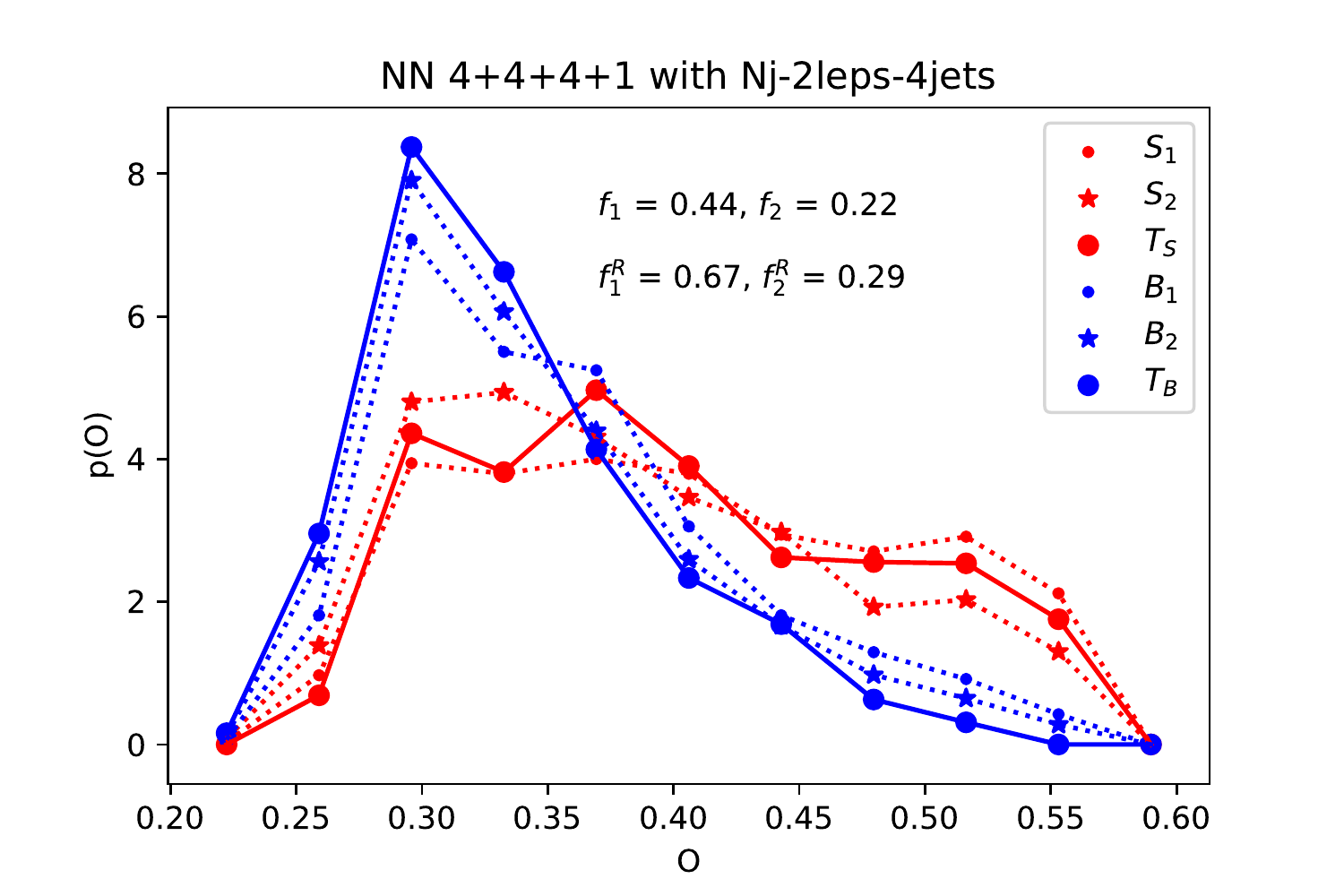}} \\\hspace{3mm}
	\caption{Neural Network outputs with three layers of four neurons preceding a single neuron layer fed with three different set of observables with the label representing whether they belong to $M_1$ or $M_2$. Left panels: Event distribution according to the trained NN output for the events in $M_1$ and $M_2$.  Right panel: Reconstructed signal ($S_R$) and background ($B_R$) distributions and underlying truth-level signal and background distributions for $M_1$ and $M_2$ samples ($S_1$, $S_2$, $B_1$ and $B_2$).  See discussion in text.}
\label{fig:NN4441}
\end{center}
\end{figure}

\begin{figure}[h!]
\begin{center}
\subfloat[]{\includegraphics[width=0.45\textwidth]{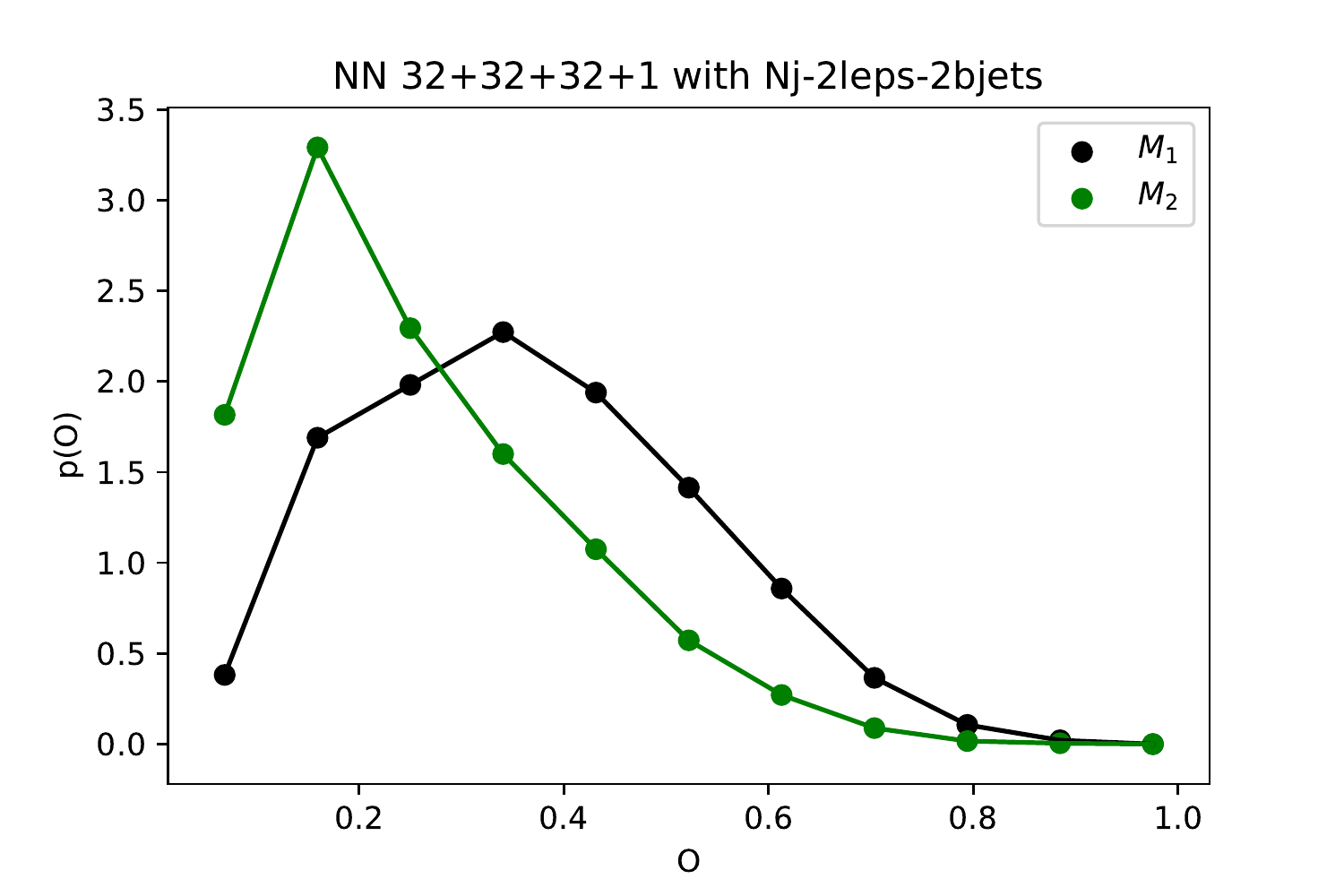}}\hspace{3mm}
\subfloat[]{\includegraphics[width=0.45\textwidth]{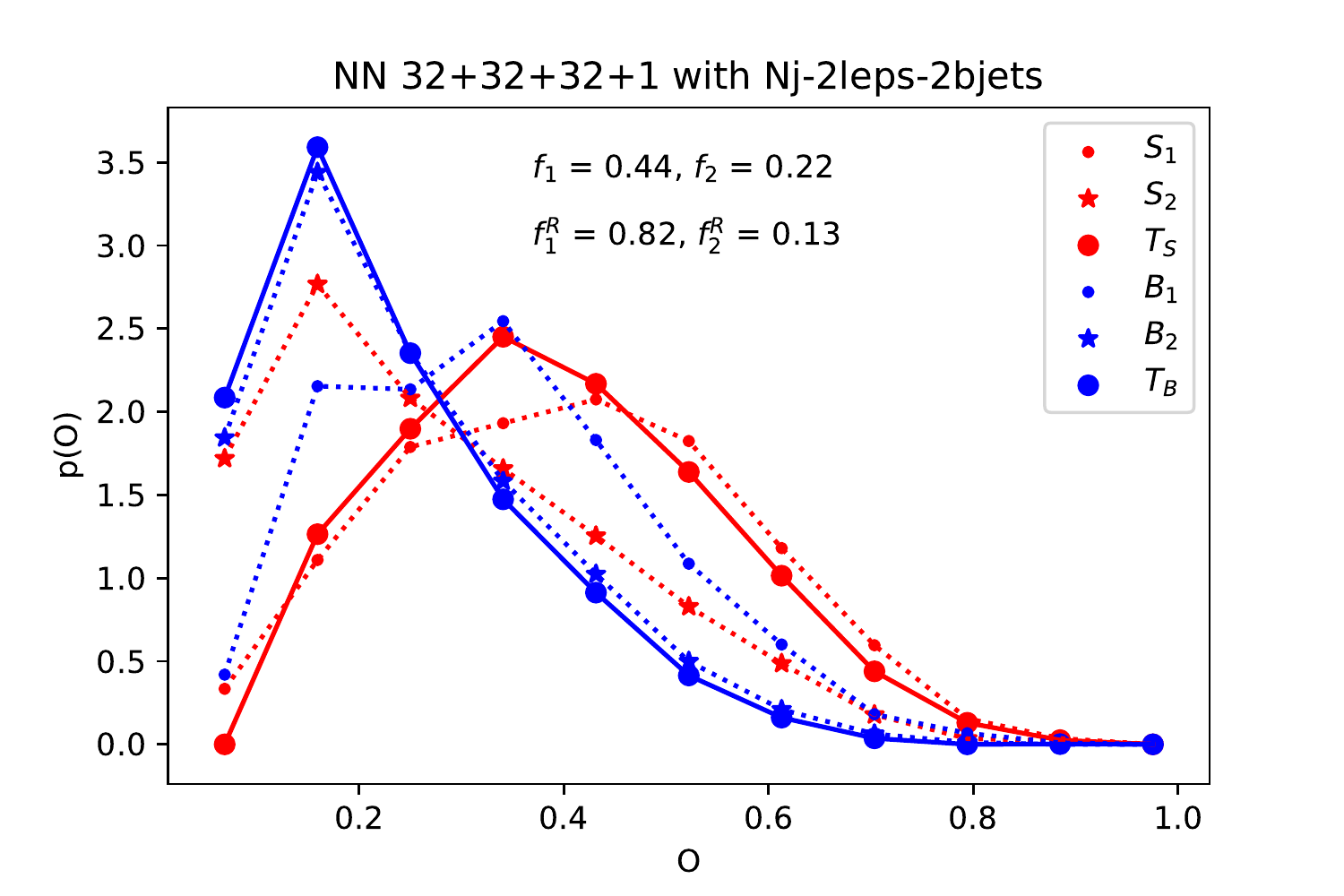}} \\\hspace{3mm}
\subfloat[]{\includegraphics[width=0.45\textwidth]{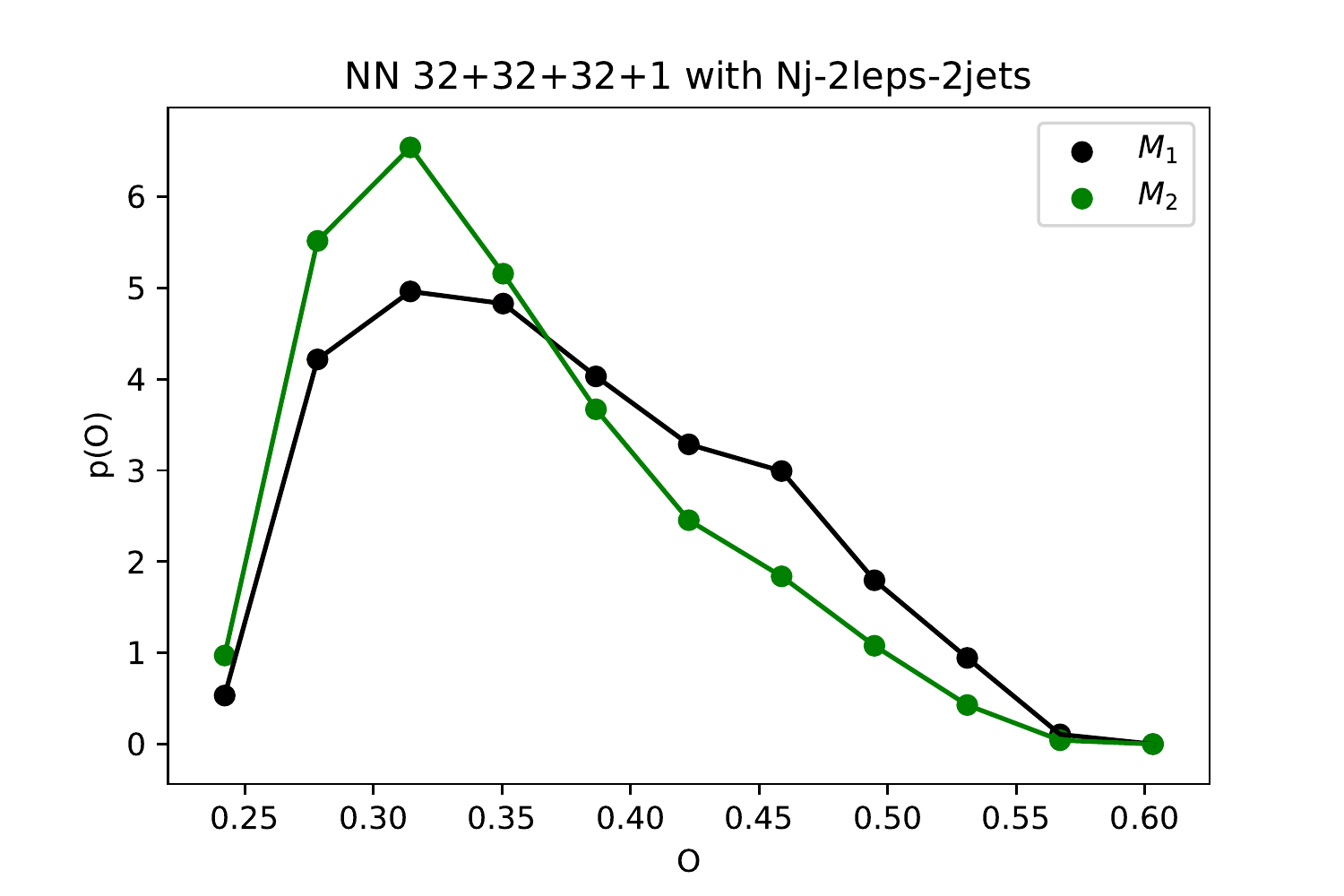}}\hspace{3mm}
\subfloat[]{\includegraphics[width=0.45\textwidth]{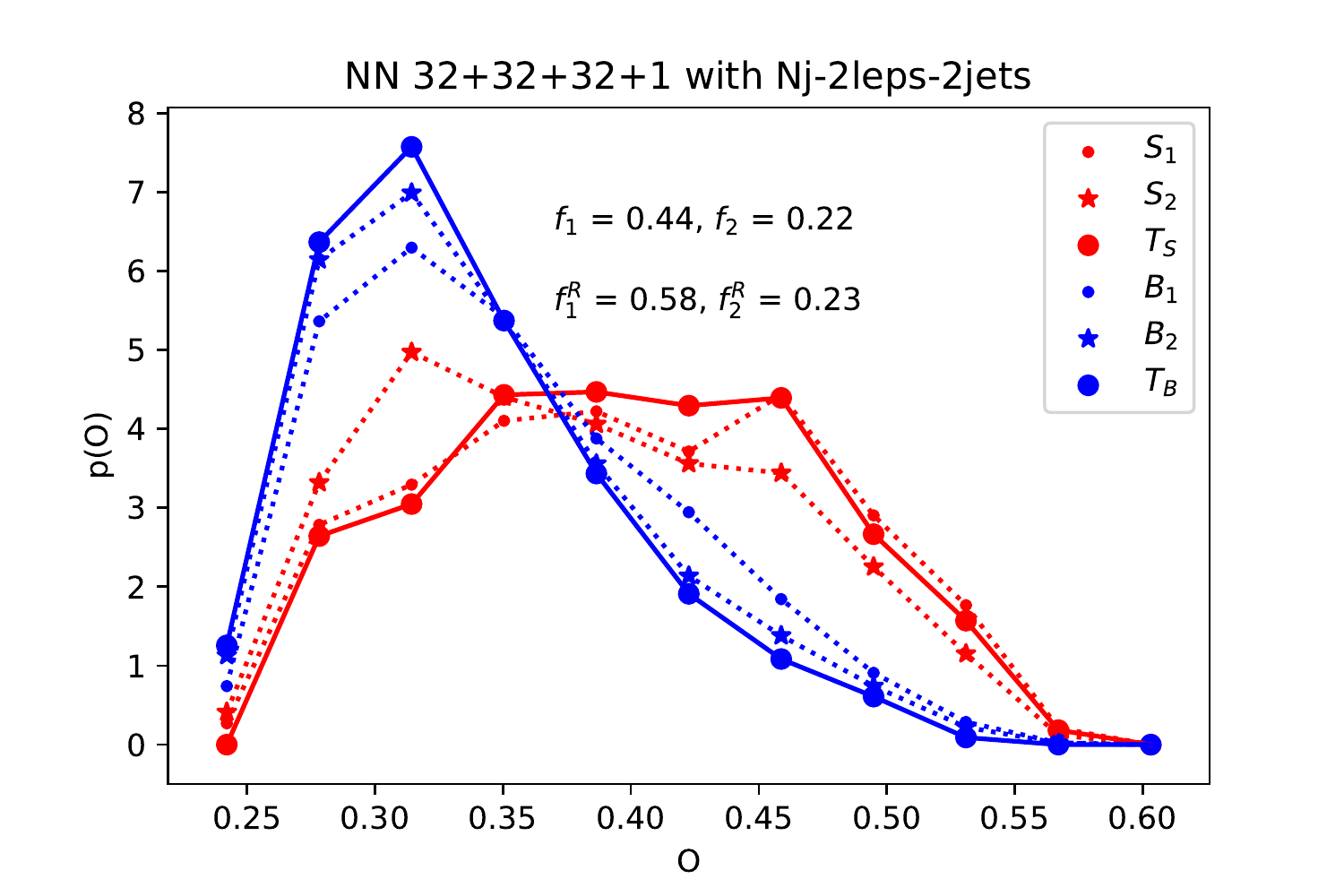}} \\\hspace{3mm}
\subfloat[]{\includegraphics[width=0.45\textwidth]{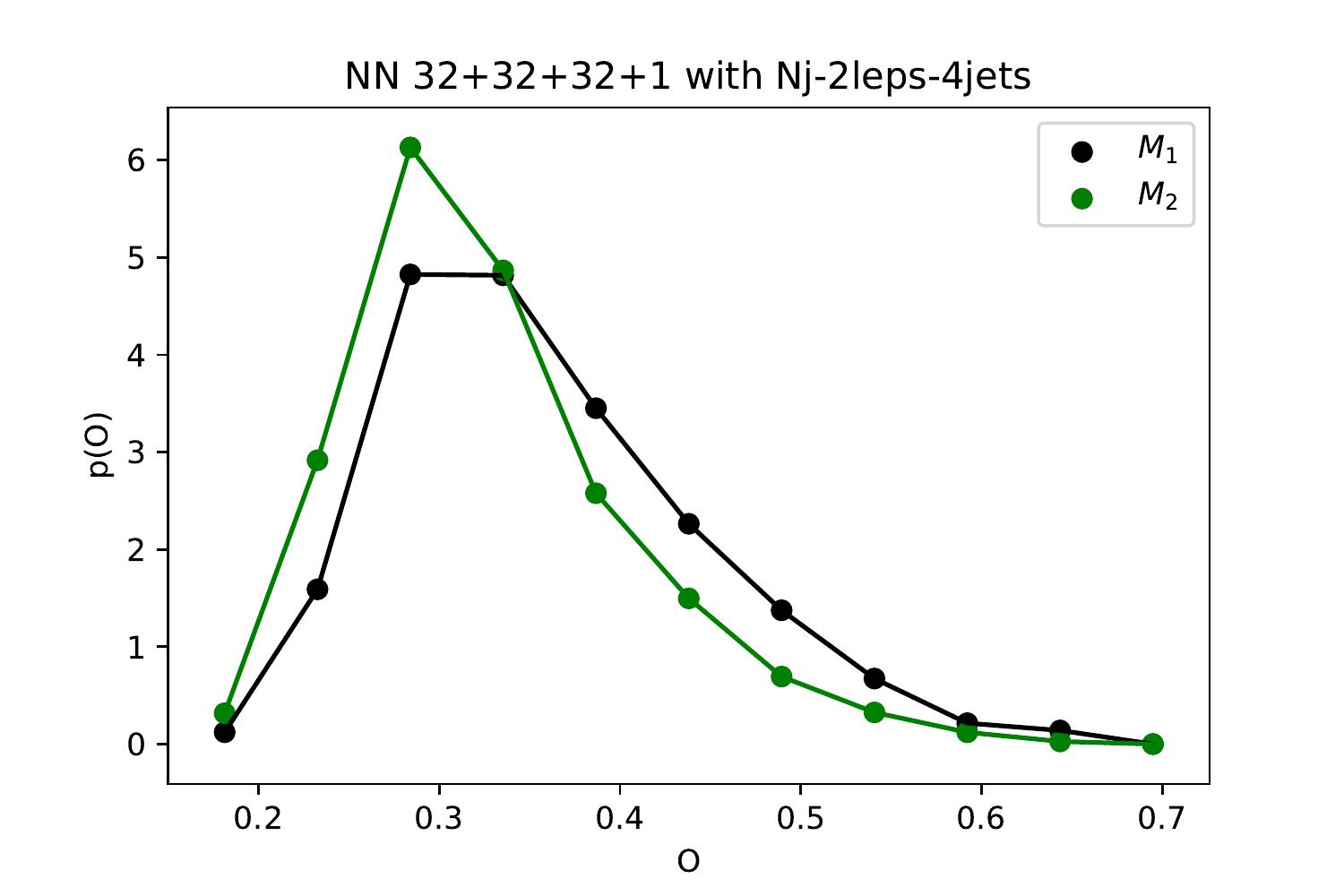}}\hspace{3mm}
\subfloat[]{\includegraphics[width=0.45\textwidth]{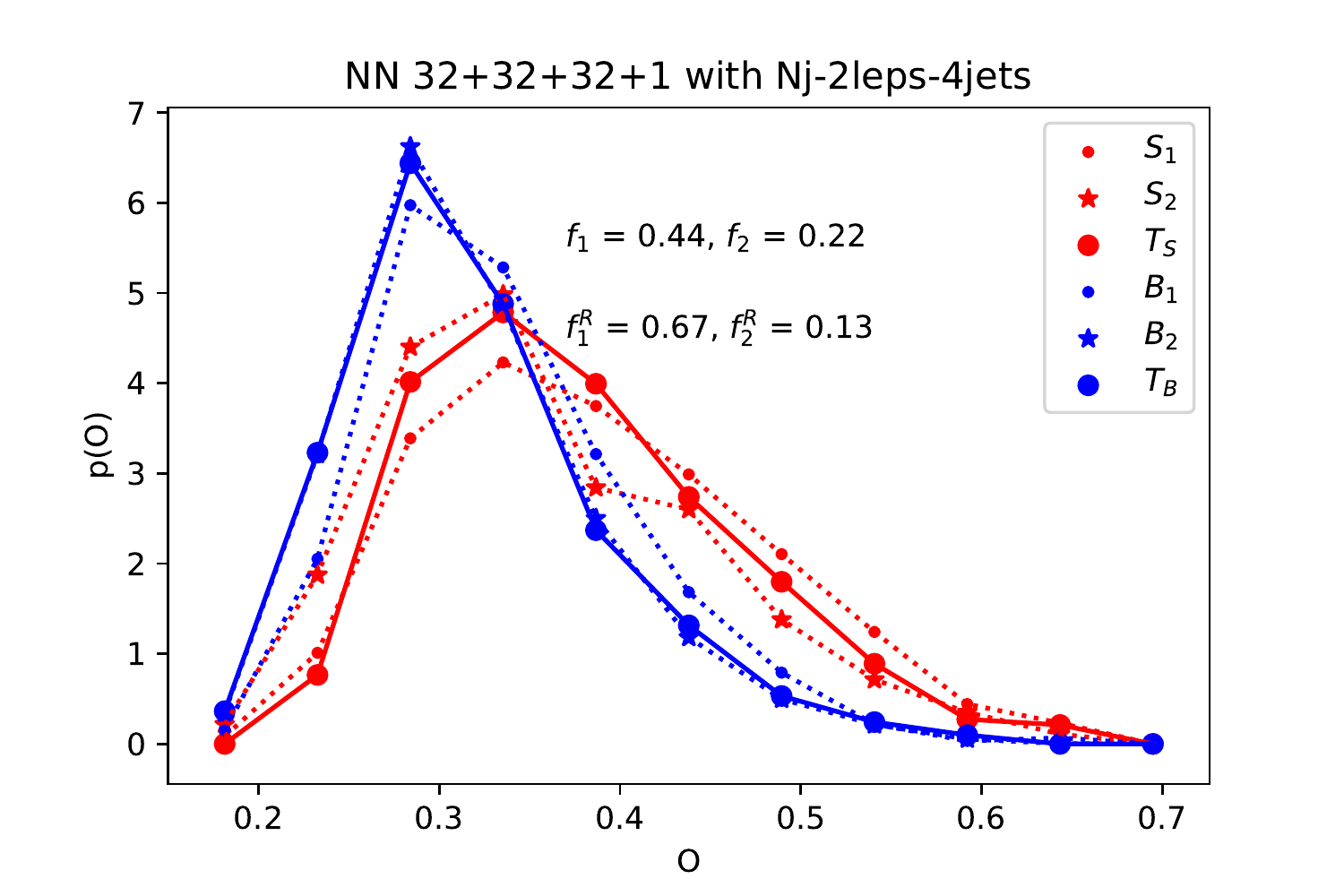}} \\\hspace{3mm}
	\caption{Idem as Fig.~\ref{fig:NN4441}, but for a more complex 32+32+32+1 NN.}
\label{fig:NN3232321}
\end{center}
\end{figure}

\begin{figure}[h!]
\begin{center}
\subfloat[]{\includegraphics[width=0.45\textwidth]{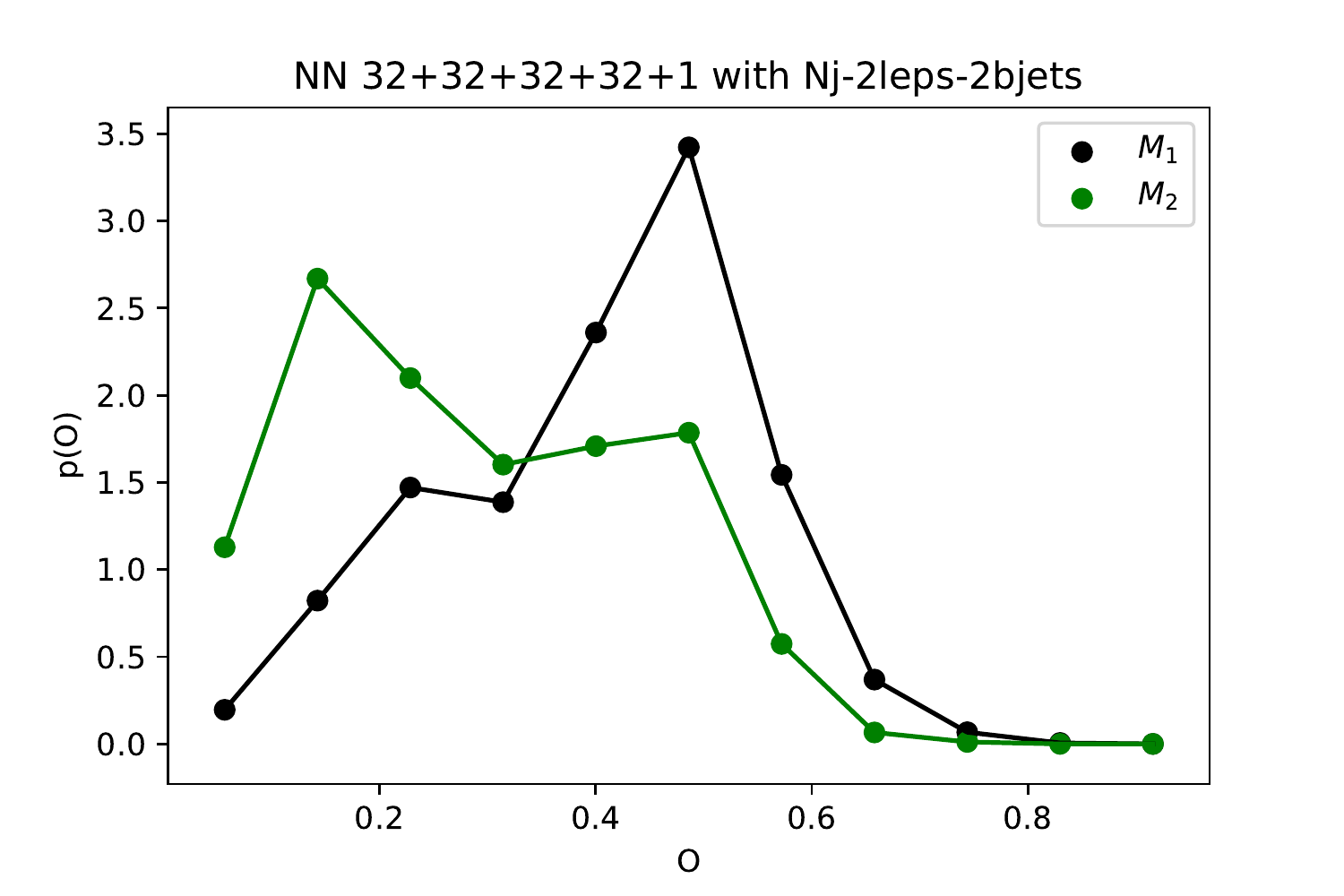}}\hspace{3mm}
\subfloat[]{\includegraphics[width=0.45\textwidth]{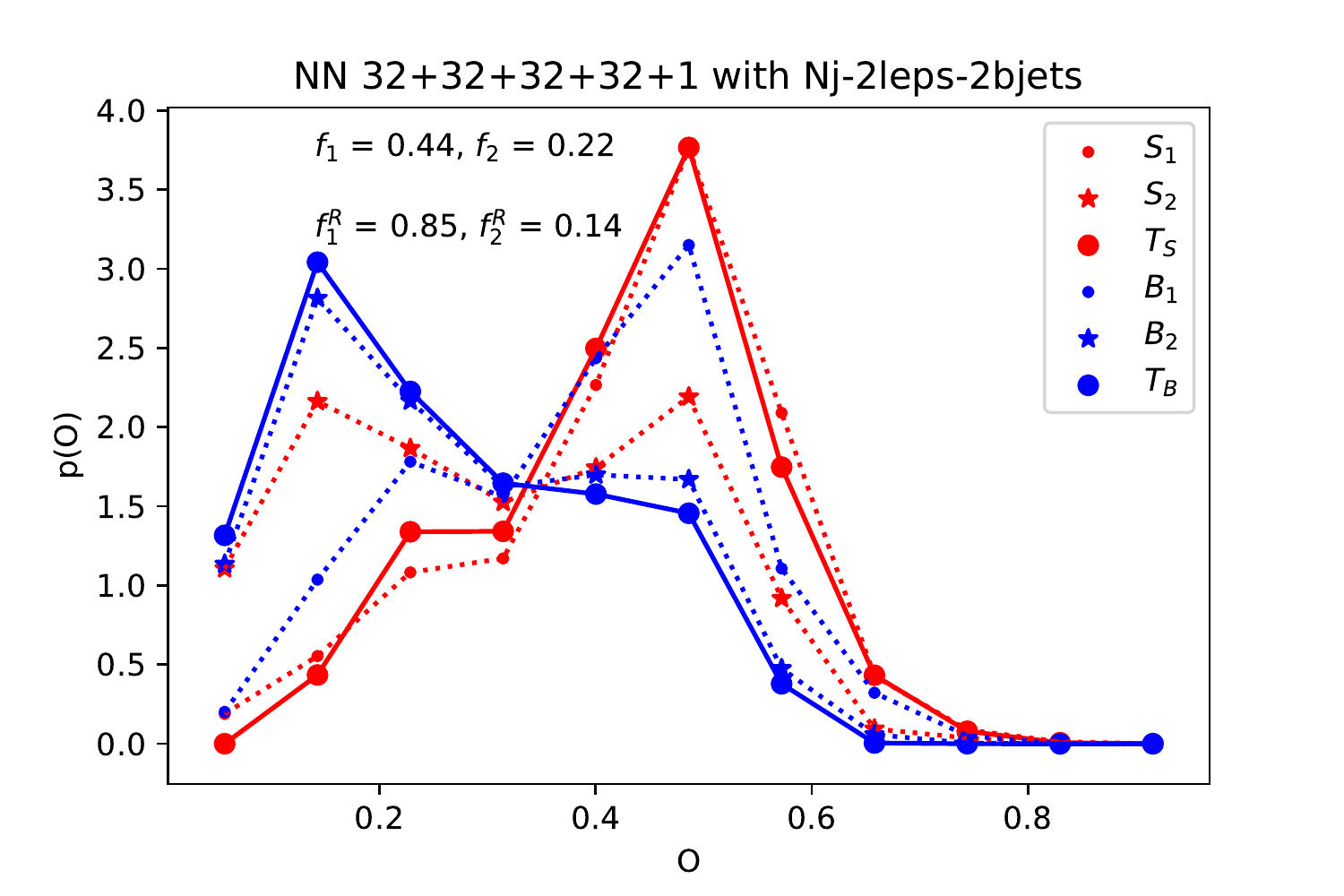}} \\\hspace{3mm}
\subfloat[]{\includegraphics[width=0.45\textwidth]{plots/NN13_m1m2.pdf}}\hspace{3mm}
\subfloat[]{\includegraphics[width=0.45\textwidth]{plots/NN13_sb.pdf}} \\\hspace{3mm}
\subfloat[]{\includegraphics[width=0.45\textwidth]{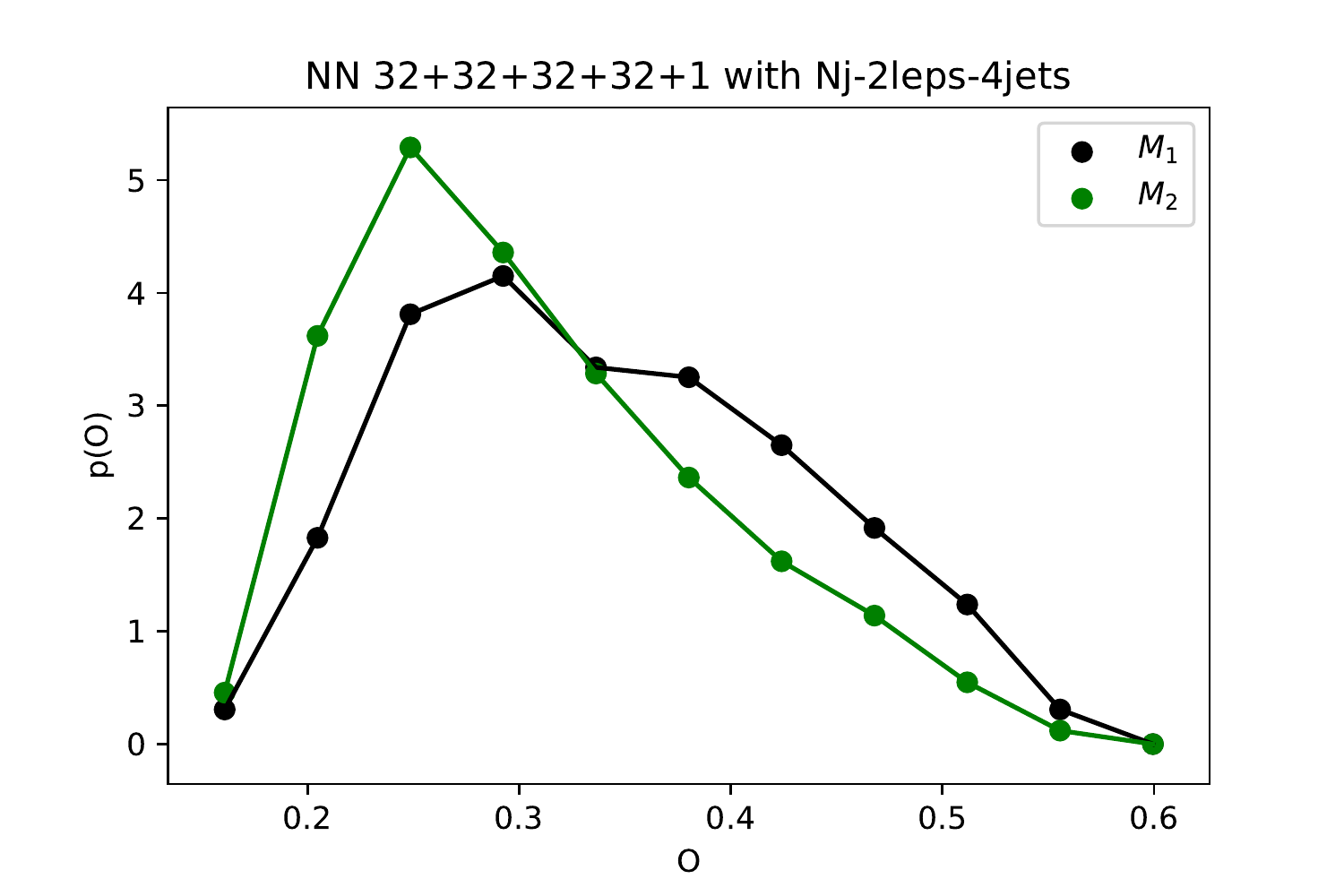}}\hspace{3mm}
\subfloat[]{\includegraphics[width=0.45\textwidth]{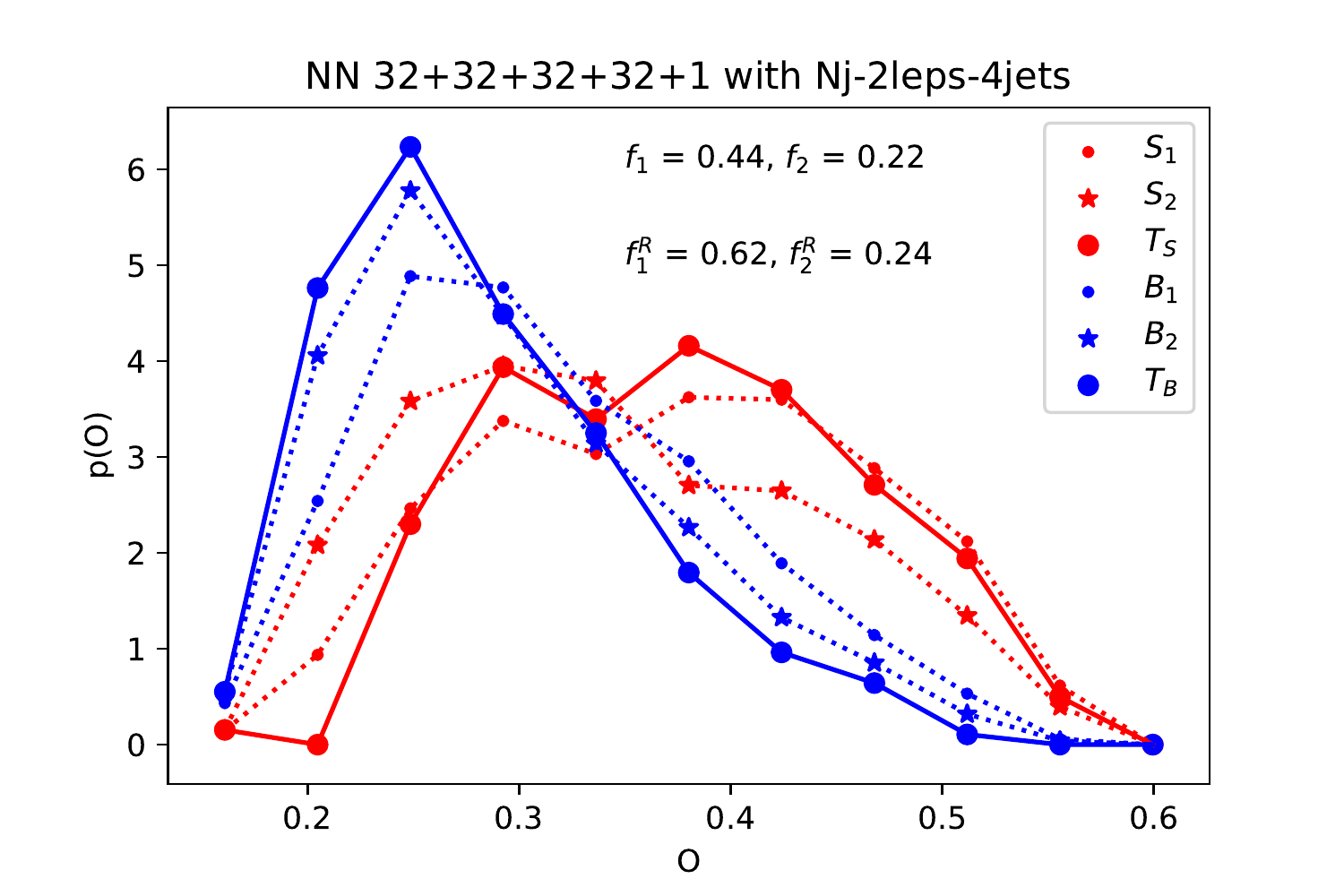}} \\\hspace{3mm}
\caption{Idem as Fig.~\ref{fig:NN4441}, but for a more complex and deeper 32+32+32+32+1 NN.}
\label{fig:NN323232321}
\end{center}
\end{figure}

Along Figs.~\ref{fig:NN4441}, \ref{fig:NN3232321} and \ref{fig:NN323232321} we show three simple architectures with three different choices of observables in each figure. We plot the NN output for $M_1$ and $M_2$ and the result of using the demixing algorithm on this output to recover the underlying distributions and the signal fractions in each mixture. As a general result, we find that using b-jets in feeding the NN brings correlation between the demixing and the samples and then, even if the discrimination may be efficient, the algorithm fails to reconstruct the underlying distributions since they do not accomplish H2.  If we instead feed the NN with the leading jets --regardless of whether they are b-tagged or not-- then such a correlation is suppressed and the reconstruction of the underlying distributions is better.  We also find that using less jets proves to be slightly better.   In Fig.~\ref{fig:NN4441} we investigate NN with 3 layers of 4 nodes each and a final sigmoid node, which we refer to it as ``4+4+4+1''.  We find that the NN is too simple and the reconstructed and underlying topics consist of tacking lines.   If Fig.~\ref{fig:NN3232321} we add nodes and use a 32+32+32+1 NN to obtain smoother curves for reconstructed and underlying topics.  Adding still one more layer with 32 nodes (Fig.~\ref{fig:NN323232321}) brings still more smoothness and a good agreement in the reconstructed fractions.    

We find the best set up by using a 32+32+32+32+1 NN fed with $N_j$, the two leading leptons and the two leading jets, Fig.~\ref{fig:NN323232321}c and Fig.~\ref{fig:NN323232321}d, which we also display in Fig.~\ref{fig:ml}.   Further exploration along these lines could bring better and more solid results for the main purpose of this work.

\FloatBarrier 

\bibliographystyle{JHEP}
\bibliography{biblio}
\end{document}